\PassOptionsToPackage{hypertexnames=false,hyperfootnotes=false}{hyperref}
\documentclass[fleqn,usenatbib]{mnras}

\usepackage{newtxtext,newtxmath}

\usepackage[T1]{fontenc}

\DeclareRobustCommand{\VAN}[3]{#2}
\let\VANthebibliography\thebibliography
\def\thebibliography{\DeclareRobustCommand{\VAN}[3]{##3}\VANthebibliography}

\usepackage{graphicx}	
\usepackage{amsmath}	
\usepackage{multirow}

\usepackage{orcidlink}

\AtBeginDocument{%
  \DeclareFontShape{OML}{cmm}{b}{it}{%
    <-6>cmmib5<6-7>cmmib6<7-8>cmmib7<8-9>cmmib8<9-10>cmmib9<10->cmmib10%
  }{}%
}

\newcommand{\aref}[1]{\hyperref[#1]{Appendix~\ref{#1}}}

\defcitealias{ReinaCamposKruijssen2017}{RC\&K}

\newcommand{\slug}{\texttt{slug}}
\newcommand{\cfour}{\texttt{c-4}}
\newcommand{\ncltotal}{8276 }
\newcommand{\ngalaxy}{12 }

\title[Environmental dependence of cluster demographics]{The Environmental Dependence of Star Cluster Demographics}

\author[Jianling Tang et al.]{
Jianling Tang,$^{1}$\orcidlink{0000-0001-5752-3847}\thanks{E-mail: janett.jianling@gmail.com}
Kathryn Grasha$^{1}$\orcidlink{0000-0002-3247-5321}
Mark R. Krumholz$^{1}$\orcidlink{0000-0003-3893-854X}
and Tomasz Różański$^{1}$\orcidlink{0000-0002-5819-3023}
\\
$^{1}$Research School of Astronomy and Astrophysics, Australian National University, Canberra, ACT 2611, Australia\\
}

\date{Accepted XXX. Received YYY; in original form ZZZ}

\pubyear{\the\year{}}

\begin{document}
\label{firstpage}
\pagerange{\pageref{firstpage}--\pageref{lastpage}}
\maketitle

\begin{abstract}
Both the star cluster mass function and the lifetimes of clusters may vary with galactic environment, but measuring this variation is challenging because in observational surveys real features of cluster demographics are invariably entangled with catalogue incompleteness. Here we analyse $\approx 8300$ star clusters in \ngalaxy galaxies drawn from the LEGUS survey using the \slug~Bayesian forward modelling framework coupled to our new \cfour~neural network-based completeness estimator, which allows us to incorporate realistic catalogue-inclusion probabilities directly into the likelihood and compensate for these biases. We show that this approach allows us to fit observed cluster luminosity functions with excellent fidelity at both galactic and sub-galactic scales.  We find that mass function slopes are relatively universal and broadly consistent with a power law $M^{-2}$ form, but high mass truncations vary by orders of magnitude both between and within galaxies. Our fits also strongly favour models where cluster disruption is mass-independent, but the time at which disruption begins again shows wide environmental variations. Our results demonstrate that young cluster demographics are environmentally dependent, with the clearest signal appearing at the upper end of the cluster mass function, but that these variations are not well-explained by any of the models currently in the literature, and do not correlate straightforwardly with properties such as star formation rate per unit area or strength of shear.
\end{abstract}




\section{Introduction}
\label{sec:intro}

Star clusters are fundamental products of star formation and sensitive tracers of the environments in which that process occurs \citep{2003Lada,2010Zwart}. Their initial mass distributions may depend on the gas surface density, pressure, shear, and similar properties of their formation environment \citep{2010Goddard,Kruijssen2012,Krumholz2019Review}, and their subsequent evolution is shaped by stellar mass loss, two-body relaxation, tidal shocks, encounters with giant molecular clouds, and the galactic tidal field \citep{Fall2001,Lamers2005,Gieles2006GMC,Kruijssen2011,2018Pfeffer}. Cluster mass and age distributions therefore provide observational probes for all these phenomena, and offer us the opportunity to measure how the balance between them changes in different galactic environments \citep{Larsen2002,Adamo2015,Johnson2016,Grasha2017a,Grasha2017b,Messa2018,Menon2021,Wainer2022}.

A central question in star-cluster demographics is whether the young-cluster mass function is a pure power law or contains a finite high-mass truncation. Young-cluster mass functions are commonly described either by a power law with a slope near $-2$ \citep{ZhangFall1999,FallChandarWhitmore2005, Mok2019, Mok2020}, consistent with a scale-free hierarchy of star-forming structures \citep{ElmegreenFalgarone1996,Elmegreen2006NGC628}, or by a Schechter-like function with a finite high-mass truncation \citep{Gieles2006,Larsen2009,Adamo2015,Adamo2017,Johnson2017,Wainer2022} -- see \citet{Krumholz2019Review} for a recent review. The characteristic truncation mass may depend on gas pressure, star formation-rate surface density, or the maximum mass scale of gravitationally unstable clouds \citep{Johnson2017,ReinaCamposKruijssen2017}. If there is indeed a truncation mass, it could in principle vary both between galaxies and within individual galaxies. Firmly establishing the existence of a truncation mass and measuring its variation with environment would represent a significant advance, and provide a vital clue for theories of star formation.

A second, equally important question concerns the distribution of star cluster ages. If clusters do not disrupt after formation, then in a galaxy forming stars at a steady rate one would expect to find equal numbers of clusters per unit age. However, most observations find that the number of clusters declines with age, indicating the action of a cluster disruption mechanism. However, the functional form of the decline with age is uncertain, with some authors favouring self-similar, mass-independent powerlaw-like behaviour \citep[e.g.,][]{FallChandarWhitmore2005, Fall09a, Chandar2010} and others finding more complex relationships between the mass and age distributions, and variations in this relationship with galactic environment \citep[e.g.,][]{Gieles07a, Gieles2009, Adamo2017} -- again, see \citet{Krumholz2019Review} for a review. As with the cluster mass function, firmly establishing the functional form of the cluster age distribution would have important implications for theories of star cluster formation and evolution, for example helping distinguish between models where cluster dissolution is mostly driven by local or internal processes such as gas clearing and stellar mass loss versus those where it is driven by external perturbations such as tidal shocks.

Nearby-galaxy surveys now provide the large, homogeneous cluster samples that are in principle capable of testing these ideas. In particular, the Legacy ExtraGalactic UV Survey (LEGUS) delivers a homogeneous HST-based view of star formation and cluster populations across 50 nearby star-forming galaxies spanning distances within the Local Universe \citep{Calzetti2015}. LEGUS cluster catalogues and follow-up analyses have constrained cluster demographics and evolution in spiral, dwarf, and irregular galaxies, including the shapes of cluster mass and age distributions, gas-clearing time-scales, hierarchical spatial structure, and internal structural properties such as cluster radii \citep{Grasha2015,Ryon2017,Adamo2017,Messa2018,Grasha2018,Grasha2019,Cook2019,Hannon2019,Brown2021,Cook2023}. Complementary programmes such as the Panchromatic Hubble Andromeda Treasury (PHAT; an M31 cluster survey) \citep{Johnson2015PHAT}, the Panchromatic Hubble Andromeda Treasury: Triangulum Extended Region (PHATTER; an M33 inner-disc survey) \citep{Johnson2022PHATTER}, and Physics at High Angular resolution in Nearby GalaxieS with HST (PHANGS--HST; a 38-galaxy cluster survey) \citep{Whitmore2021,Thilker2022,Maschmann2024} extend this effort in different ways. These surveys provide the statistical power needed to test whether cluster formation and evolution vary systematically with galactic environment. Interpreting those variations, however, requires careful treatment of the observational selection effects that shape cluster catalogues.

Observed cluster catalogues are selection-biased representations of the underlying cluster population. Whether a source enters the final catalogue depends on the detection depth, local background and crowding, source concentration, multi-band detection requirements, and in some cases morphological classifications \citep{Adamo2017,Whitmore2021,Thilker2022}. Artificial star cluster tests (ASTs) are widely used to characterise the resulting selection function. In this approach, one injects synthetic clusters into the images and measures the rate at which they are recovered after applying the same detection and selection procedures used to construct the observed catalogue \citep{Johnson2022PHATTER}. In many traditional demographic analyses, however, the recovery measurements from ASTs are not carried through as a continuous selection function. Instead, they are summarised by a completeness threshold such as a 90\% magnitude limit, which is then converted into a mass--age selection boundary \citep{Shabani2018}. Only clusters within the nominally complete region are retained for inference of the cluster mass and age functions \citep{Messa2018}. Other analyses similarly impose explicit mass and age cuts before fitting the cluster population \citep{Wainer2022}.

Although this procedure limits the influence of incompleteness on the retained sample, it does so at the heavy price of discarding large numbers of clusters from partially complete regions of parameter space, preferentially excluding intrinsically fainter low-mass and older clusters. For the sample analysed here, applying such cuts would remove more than half of the available clusters and thereby greatly reduce the information available to constrain cluster demographics. Moreover, this procedure encounters a fundamental limit when dealing with low-mass clusters: because clusters smaller than a $\textrm{few}\times 10^3$ M$_\odot$ have stellar populations too small to fully sample the stellar initial mass function (IMF), they do not have a deterministic mapping from cluster mass and age to magnitude that would make it possible to translate a magnitude cut into a mass or age cut \citep{Cervino04a, Fouesneau10a, DaSilva2012, Anders13a, Krumholz2015}. Consequently this approach encounters a fundamental limit to the range of star cluster mass to which it can be applied regardless of the depth of the underlying data.

To retain the information carried by clusters in partially complete regions of parameter space, and to facilitate a forward-modelling approach that circumvents the problem of the mapping from cluster mass and age to photometry being non-deterministic, \citet{Tang2026} introduced the cluster completeness correction calculator (\cfour), which uses artificial star cluster test recovery outcomes to learn a continuous, multidimensional representation of the full catalogue selection function as a function of both cluster photometric and physical properties. We combine \cfour~with the population-level inference framework introduced by \citet{Krumholz2019SLUGIV} as part of the \slug~(Stochastically Lighting Up Galaxies) software suite. In this method one starts from a proposed population-level distribution of cluster mass, age, and extinction, and then uses a stochastic synthetic cluster library to predict the corresponding distribution in multidimensional photometric space, properly incorporating photometric uncertainties, stochastic IMF sampling, and heterogeneous filter coverage. \cfour~then supplies the crucial final piece of information needed for forward-modelling: the probability that each synthetic cluster would enter the observed catalogue in the first place, thereby mapping the intrinsic photometric distribution of the population computed by \slug~to the predicted selection-biased observable distribution. The resulting photometric distribution can then be compared directly with the observed catalogue, providing a complete forward model of cluster demographics, which we can then use to adjust the parameters describing the demographics of the underlying population to reproduce the observed photometric distribution as closely as possible.


\citet{Tang2024} developed a first pilot version of this approach and applied it to NGC 628, but crucially that study relied on a rather crude estimate of completeness similar to the approaches in earlier work. The addition of the \cfour~neural network framework therefore represents a significant improvement in methodology compared to that used in \citet{Tang2024}. The central aim of this paper is to determine whether this improved method allows us to obtain firm evidence for environmentally dependent cluster formation and evolution. Moreover, rather than a pilot study of one galaxy, here we apply our forward-modelling framework to all of the galaxies in the LEGUS survey that have cluster catalogues large enough for meaningful population inference. There are \ngalaxy~such galaxies, which span a broad range of morphologies and star formation-rate surface densities. Our primary goals are to test whether the cluster mass function requires a finite high-mass truncation, whether the cluster age distribution depends on cluster mass or not, and whether the answers to either of these questions, and the parameters describing these answers, vary with galactic environment on both whole-galaxy and sub-galactic scales. 

This paper is organised as follows. \autoref{sec:data} describes the LEGUS data and cluster catalogues. \autoref{sec:methods} presents the demographic model and completeness treatment. \autoref{sec:results} reports the galaxy-wide and sub-galactic results. \autoref{sec:discussion} discusses the implications for environmentally regulated cluster formation, with secondary consideration of cluster disruption and stellar-population model systematics. \autoref{sec:summary} summarises our conclusions.

\section{Data}
\label{sec:data}
\subsection{Galaxy sample}
In this study we use a sample of \ngalaxy galaxies, listed in \autoref{tab:galaxy_properties}, selected from the Legacy ExtraGalactic UV Survey (LEGUS) Hubble Space Telescope survey. LEGUS provides near-ultraviolet to $I$-band imaging of nearby ($D < 18$~Mpc) star-forming galaxies, primarily using WFC3/UVIS filters F275W, F336W, F438W, F555W, and F814W, with existing archival ACS imaging \citep{Calzetti2015, Grasha2015,Adamo2017}. Our sample includes all the LEGUS galaxies with publicly available star cluster catalogues that include a number of clusters large enough for meaningful statistical modelling ($N>70$). Some of these galaxies have a single \textit{HST} field in LEGUS, while others are covered by multiple fields, in some cases with different combinations of filters -- see the second column of \autoref{tab:galaxy_properties}.

The selected galaxies span nearly one orders of magnitude in star formation rate surface density, include grand-design spirals, flocculent spirals, and dwarf irregulars, and cover a broad range of stellar masses, gas fractions, and dynamical environments. This diversity allows us to test whether cluster demographics vary systematically with global galaxy properties. The sample is therefore designed to maximize dynamic range in environmental conditions rather than statistical completeness.

\subsection{LEGUS star cluster catalogues}

For each of our target galaxies, we make use of the LEGUS public star cluster catalogues. There is one such catalogue per field, rather than one per galaxy, though for our analysis we will in many cases combine the catalogues from different fields in the same galaxy (see \autoref{sec:methods}). A detailed description of these catalogues, including the full catalogue construction pipeline, can be found in \citet{Calzetti2015} and \citet{Adamo2017}. These catalogues provide integrated photometry in five \textit{HST} bands, positions, morphological classifications, and SED-derived estimates of age, mass, and extinction for each identified star cluster in the field covered by LEGUS. While we use these catalogue products to define the observed cluster population, we do not adopt the catalogue-derived physical parameters in our demographic inference, instead relying on our forward modelling approach for the reasons discussed in \autoref{sec:intro}, and which we describe in detail in \autoref{sec:methods}. 



Before making use of the catalogues, we apply one correction to them: the vast majority of the star clusters in the catalogues were selected using the LEGUS automatic pipeline, but a small number, mostly ones that were slightly below the magnitude limit for the automatic catalogue, were added by hand. Our forward modelling approach relies on accurate calculation of catalogue completeness, which we can obtain for the automatic catalogue but not for the hand-selected additions, so it is important to remove these hand-added clusters from the sample. Unfortunately the publicly available LEGUS catalogues do not distinguish between the clusters that were found by the automatic pipelines and those added manually, so we identify the manual clusters for removal by using our \cfour~modelling tool, which uses a neural network trained on artificial star cluster tests to produce a high-fidelity model of the LEGUS automatic pipeline -- see \citet{Tang2026} and \autoref{subsec:completeness} for details. This tool can predict the probability $\widehat{P}_{\rm obs}$ that a given cluster will be recovered by the LEGUS automatic pipeline either as a function of its physical properties (mass, age, etc.) or its photometric magnitudes. While we will use the former capability in the main part of this work, here we use the latter method to identify clusters in the public catalogues whose reported photometric magnitudes indicate that they could not have been recovered by the automatic pipeline (i.e., \cfour~predicts $\widehat{P}_{\rm obs} = 0$), and remove them. This cut removes 114 clusters across all the catalogues we use, a small fraction (1.36\%) of the total catalogue size. Almost all of the clusters found by this procedure have V-band magnitudes dimmer than the magnitude cut-off of the LEGUS automatic catalogue, confirming that our method is identifying genuine hand-additions that could not have been produced by the automatic catalogue construction procedure.

Applying this cut, we arrive at a final sample of \ncltotal clusters; the number of clusters per galaxy is listed in \autoref{tab:galaxy_properties}. Our sample does not discriminate based on cluster morphology, and includes clusters assigned morphological types 1, 2, and 3 in the LEGUS catalogues (see \citealt{Adamo2017} for details).

\begin{table*}
\centering

\small
\setlength{\tabcolsep}{3pt}
\begin{tabular}{lp{8.0cm}ccccr}
\hline
Galaxy & Filters and cameras & Morph. & $D$
& $\log \Sigma_{\rm SFR}$ & $N_{\rm cl}$ \\
& & & (Mpc)
& (${\rm M_\odot\,yr^{-1}\,kpc^{-2}}$)
&  & \\
\hline

NGC 0628 &
Central (C): WFC3/UVIS
(F275W, F336W, F555W), ACS/WFC
(F435W, F814W);
\newline
Edge (E): WFC3/UVIS
(F275W, F336W), ACS/WFC
(F435W, F555W, F814W)
& SAc & 9.8  & $-2.36$ & 1236 \\

NGC 1313 &
WFC3/UVIS (F275W, F336W),
ACS/WFC (F435W, F555W, F814W)
& SBd & 4.3  & $-1.96$ & 727 \\

NGC 1566 &
WFC3/UVIS
(F275W, F336W, F438W, F555W, F814W)
& SABbc & 17.7  & $-2.41$ & 1513 \\

NGC 3344 &
WFC3/UVIS
(F275W, F336W, F438W, F555W, F814W)
& SABbc & 9.8 & $-2.57$ & 383 \\

NGC 3627 &
WFC3/UVIS (F275W, F336W, F438W, F555W, F814W)
& SABb & 11.3  & $-1.64$ & 739 \\

NGC 3738 &
WFC3/UVIS (F275W, F336W, F438W),
ACS/WFC (F606W, F814W)
& Im & 5.1 & $-2.19$ & 206 \\

NGC 4449 &
WFC3/UVIS (F275W, F336W),
ACS/WFC (F435W, F555W, F814W)
& IBm & 4.0 & $-1.64$ & 423 \\

NGC 5194-NGC 5195 &
WFC3/UVIS (F275W, F336W),
ACS/WFC (F435W, F555W, F814W)
& SAbc & 8.6 & $-1.77$ & 1228 \\

NGC 5253 &
WFC3/UVIS (F275W, F336W),
ACS/WFC (F435W, F555W, F814W)
& Im & 3.3 & $-2.26$ & 70 \\

NGC 5457 &
WFC3/UVIS (F275W, F336W),
ACS/WFC (F435W, F555W, F814W)
& SABcd & 6.7 & $-2.36$ & 1144 \\

NGC 6503 &
WFC3/UVIS
(F275W, F336W, F438W, F555W, F814W)
& SAcd & 6.3 & $-2.60$ & 293 \\

NGC 7793 &
East (E): WFC3/UVIS
(F275W, F336W, F438W, F555W, F814W);
\newline
West (W): WFC3/UVIS
(F275W, F336W, F438W),
ACS/WFC (F555W, F814W)
& SAd & 3.6 & $-2.17$ & 314 \\

\hline
\end{tabular}

\vspace{0.5em}
\begin{minipage}{0.99\textwidth}
\footnotesize
\caption{Galaxy-wide properties, LEGUS filter--camera combinations, and
final cluster sample sizes. The filter--camera combinations are adopted from
Table~2 of \citet{Calzetti2015}. WFC3/UVIS denotes the
ultraviolet--visible channel of the \textit{Hubble Space Telescope}
Wide Field Camera~3, whereas ACS/WFC denotes the Wide Field Channel
of the Advanced Camera for Surveys. Morphological classifications, distances, and star formation rate
surface densities are adopted from Table~1 of \citet{Menon2021}.
The logarithmic star formation rate surface density is defined as
$\log \Sigma_{\rm SFR} \equiv
\log_{10}[\Sigma_{\rm SFR}/
({\rm M_\odot\,yr^{-1}\,kpc^{-2}})]$. The cluster counts $N_{\rm cl}$ correspond to the final catalogue size of each galaxy after removing sources that do not satisfy the LEGUS selection criteria.
\label{tab:galaxy_properties}
}
\end{minipage}
\end{table*}

\subsection{Ancillary data}
\label{sec:environmental_measurements}

In the analysis that follows we will be interested in how star cluster demographics vary with other galactic properties, either galaxy-wide or locally. To enable this analysis we add two pieces of ancillary data: star formation surface densities $\Sigma_\mathrm{SFR}$ and rotation curves $v_\mathrm{circ}(r)$.

We characterize the star formation intensity of each galaxy using the star formation rate surface density, $\Sigma_{\rm SFR}$. In \autoref{tab:galaxy_properties}, we report only a representative central value for each galaxy.\footnote{Spatial variation in $\Sigma_{\rm SFR}$ is represented by the 16th--84th percentile range across valid 1.5-kpc apertures for NGC~628, NGC~1566, and NGC~7793 \citep{Sun2022,Sun2023}, and by the narrowest 68 percent interpercentile range reported for NGC~5194--NGC~5195 \citep{Wainer2022}.} For NGC~628, NGC~1566, and NGC~7793, we adopt the median of the local $\Sigma_{\rm SFR}$ measurements across all valid 1.5-kpc apertures in the PHANGS MegaTables \citep{Sun2022,Sun2023}. For NGC~5194--NGC~5195, we adopt the central value reported by \citet{Wainer2022}. For the remaining galaxies, we adopt the values listed in Table~1 of \citet{Menon2021}, which are based on \citet{Calzetti2015}. These values are averaged over either the $R_{25}$ disc or the LEGUS field of view. 

We use the galaxy rotation curves, $v_{\rm circ}(r)$, to characterise the local dynamical environment through their logarithmic slope, $\beta \equiv {\rm d}\ln v_{\rm circ}/{\rm d}\ln r$. Profiles of $\beta$ versus $r$ for three of our sample galaxies are available in \citet{SuwannajakTanLeroy2014}: NGC~5457, NGC~5194--NGC~5195, and NGC~628. Below, we will investigate the possible effects of shear on cluster demographics by dividing these galaxies at the galactocentric radius where $\beta=0.3$, following the threshold used by \citet{SuwannajakTanLeroy2014} to distinguish low- and high-shear regimes. The dividing radii are 2.4~kpc for NGC~5457 and 1.6~kpc for both NGC~5194--NGC~5195 and NGC~628. The resulting split separates an inner, higher-$\beta$ region, where the rotation curve is closer to solid-body rotation, from an outer, lower-$\beta$ region, where the rotation curve is closer to flat.

\section{Methods}
\label{sec:methods}
We infer the intrinsic cluster mass, age, and extinction distributions using the Bayesian forward-modelling framework developed in \citet{Krumholz2019SLUGIV} as part of the \slug~software suite, improved by the addition of the \cfour~neural-network completeness model of \citet{Tang2026} to account for the probability that synthetic clusters would be captured in the observed catalogues. This section summarises the elements needed for the present analysis; full algorithmic details, validation tests, and implementation choices are given in those papers.

\subsection{Bayesian inference method}

We consider a star cluster to be described by three physical parameters\footnote{In principle metallicity [Fe/H] could form a fourth physical parameter. However, prior experiments have shown that for the cluster populations of nearby star-forming galaxies, metallicity has relatively small effects on cluster inference \citep[e.g.,][]{Krumholz15c}, and so we for simplicity we ignore it.} -- mass $M$, age $T$, and extinction $A_{\mathrm{V}}$ -- and a vector of photometric magnitudes $\boldsymbol{m}$ in various filters. The relationship between the physical and photometric properties is neither unique nor deterministic; that is, two clusters with the same $(M,T,A_V)$ may nonetheless have different photometric magnitudes $\boldsymbol{m}$ due to their stellar populations including different individual stellar masses, and two clusters with different $(M,T,A_V)$ may nonetheless produce identical photometric vectors $\boldsymbol{m}$ due both to stochastic sampling of the IMF and degeneracies between parameters (e.g., the usual age-extinction degeneracy, whereby increasing age and extinction both redden a population in similar ways). Consequently, we cannot infer a unique value of the parameters $(M,T,A_V)$ from an observed set of magnitudes $\boldsymbol{m}$.

We assume that the distribution of physical properties for a star cluster population under study is described by a distribution function $f(M, T, A_V \mid \boldsymbol{\theta})$, where $\boldsymbol{\theta}$ is a vector of parameters describing the distribution; for example, below we will consider classes of model in which the cluster mass function is described by a Schechter function, in which case $\boldsymbol{\theta}$ includes the powerlaw slope $\alpha_M$ and break mass $M_\mathrm{break}$ of that Schechter function. The goal of our analysis is to infer these parameters $\boldsymbol{\theta}$ from observations of a galaxy, or a portion therefore, in which we measure $N_\mathrm{obs}$ clusters, each of which has a vector of magnitudes $\boldsymbol{m}_i$ with associated observational uncertainties $\boldsymbol{\sigma}_i$, for $i = 1 \ldots N_\mathrm{obs}$; we let $\{\boldsymbol{m}\}$ and $\{\boldsymbol{\sigma}\}$ represent this collection of measurements and uncertainties.

Note that, because LEGUS observations of different fields used different filter combinations, and some catalogues lack measurements in all five bands for all clusters (e.g., due to offsets in the exact coverage footprint of the different filters used for that field), the set of magnitude measurements available may not be the same for all clusters. To accommodate this, we group the observations in sets that share the same set of filters, which we denote $\{\boldsymbol{m}\}_\mathcal{F}$. That is all $N_{\mathrm{obs},\mathcal{F}}$ of the observed clusters in filter group $\mathcal{F}$ (out of $N_\mathcal{F}$ groups) share the same set of filters, and $\sum_{\mathcal{F}=1}^{N_\mathcal{F}} N_\mathrm{obs,\mathcal{F}} = N_\mathrm{obs}$. Per the usual Bayesian approach, we write the formal solution to this inference problem in terms of Bayes' Theorem, which allows us to write the posterior probability distribution for $\boldsymbol{\theta}$ under this set of measurements as
\begin{equation}
    p(\boldsymbol{\theta}\mid \{\boldsymbol{m}\}, \{\boldsymbol{\sigma}\}) \propto \mathcal{L}(\{\boldsymbol{m}\} \mid \boldsymbol{\theta}, \{\boldsymbol{\sigma}\})\; p_\mathrm{prior}(\boldsymbol{\theta}),
    \label{eq:posterior}
\end{equation}
where $p_\mathrm{prior}(\boldsymbol{\theta})$ is our prior probability distribution and $\mathcal{L}(\{\boldsymbol{m}\} \mid \boldsymbol{\theta}, \{\boldsymbol{\sigma}\})$ is the likelihood function describing the probability density of the measured data given the observational uncertainties and the parameters that we seek to infer.

Following \citet{Krumholz2019SLUGIV} and \citet{Tang2024}, we evaluate the likelihood function by constructing a library of synthetic clusters using the \slug~stochastic stellar population synthesis code. Each synthetic cluster has a vector of physical parameters $(M_j, T_j, A_{V,j})$ and a vector of magnitudes $\boldsymbol{m}_j$, which we compute for all the filters found in the observations to which we are comparing. For all the inferences carried out in this paper, we use a library of $10^7$ synthetic star clusters generated following the same approach as \citet{Tang2024}: we assume a \citet{Chabrier05a} IMF, use \slug's ``starburst99'' spectral-synthesis mode \citep{Leitherer99a, Vazquez05a}, and adopt a Milky Way extinction curve \citep{Landini84a, Fitzpatrick99a}; see \citet{Krumholz2015} for a full description of how these options are implemented. Only the treatment of nebular emission differs (slightly) from that in \citet{Tang2024}; here we instead follow the improved method described in the appendix of \citet{Tang2026}. The
library cluster masses are drawn
from $p(M)\propto M^{-1}$ over $10^2 \le M/M_\odot < 10^5$ and from
$p(M)\propto M^{-2}$ over $10^5\le M/M_\odot \le 10^7$; ages are drawn
uniformly in $\log(T)$ from $10^5~{\rm yr}$ to $1.5\times10^{10}~{\rm yr}$; and
extinctions are drawn uniformly over $0 \le A_V \le 3~{\rm mag}$. We compute
integrated cluster photometry in the Vega system for the \textit{HST}
filters listed in \autoref{tab:galaxy_properties} including nebular emission and extinction. 

\citet{Krumholz2019SLUGIV} and \citet{Tang2024} show that the likelihood function can then be computed from this library as
\begin{eqnarray}
    \lefteqn{
    \mathcal{L}(\{\boldsymbol{m}\} \mid \boldsymbol{\theta}, \{\boldsymbol{\sigma}\}) \propto
    } \nonumber \\
    & &
    \prod_{\mathcal{F}=1}^{N_\mathcal{F}}\left\{
    \prod_{i=1}^{N_{\mathrm{obs},\mathcal{F}}} \mathcal{A}_\mathcal{F}(\boldsymbol{\theta}) \left[\sum_{j=1}^{N_\mathrm{lib}} w_{j,\mathcal{F}}(\boldsymbol{\theta}) \mathcal{N}(\boldsymbol{m}_i - \boldsymbol{m}_j, \boldsymbol{h}_i')\right]\right\}.
    \label{eq:likelihood}
\end{eqnarray}
Here the outermost product runs over all the filter sets available, the inner product runs over all the observed clusters for that filter set, and the inner sum runs over all library clusters. The quantity $\mathcal{N}(\boldsymbol{x}, \boldsymbol{h}'_i)$ is the usual multidimensional Gaussian function (with the number of dimensions depending on the number of magnitudes available in a given filter set) with central value $\boldsymbol{x}$ and dispersion $\boldsymbol{h}'_i$, and the bandwidth $\boldsymbol{h}_i' = (h^2 + \boldsymbol{\sigma}_i^2)^{1/2}$ that we use is a quadrature sum of an intrinsic bandwidth $h = 0.25$ mag that is set based on the sampling density of our library and the observational uncertainties $\boldsymbol{\sigma}_i$ of the $i$th cluster in each photometric band.

The parameters $w_{j,\mathcal{F}}(\boldsymbol{\theta})$ are weights applied to each cluster in the synthetic library, which combine the intrinsic frequency of a cluster with those properties in the population and the probability that such a cluster would be included in the observed catalogue after accounting for observational incompleteness; 
the normalisation factor
\begin{equation}
\mathcal{A}_\mathcal{F}(\boldsymbol{\theta})
=
\left[
\sum_{j=1}^{N_{\rm lib}}
w_{j,\mathcal{F}}(\boldsymbol{\theta})
\right]^{-1}.
\label{eq:filter_set_normalization}
\end{equation}
is the inverse sum of these weights. Formally, we write the weight of each library cluster for a given filter set as
\begin{equation}
    w_{j,\mathcal{F}}(\boldsymbol{\theta}) =
    \widehat{P}_{\rm obs,\mathcal{F}}(\boldsymbol{m}_j)
    \frac{f(M_j,T_j,A_{{\rm V},j}\mid\boldsymbol{\theta})}
         {p_{\rm lib}(M_j,T_j,A_{{\rm V},j})},
    \label{eq:library_weights}
\end{equation}
where $p_{\rm lib}$ is the library sampling density (i.e., the probability distribution from which we drew values of $(M,T,A_V)$ when making the library, as described above) and $\widehat{P}_{\rm obs}$ is the catalogue-inclusion probability. We compute $\widehat{P}_{\rm obs}$ using our newly-developed \cfour~completeness modelling software, which we describe in the next section. We carry out numerical evaluation of the likelihood function using \texttt{cluster\_slug}, which implements a fast (order $N_\mathrm{obs} \ln N_\mathrm{lib}$) algorithm to evaluate \autoref{eq:likelihood}.

\subsection{Completeness calculation}
\label{subsec:completeness}


We evaluate $\widehat{P}_{\rm obs}$ using \cfour, Cluster Completeness Correction Calculator, a neural-network-based completeness model introduced by \citet{Tang2026}. Rather than imposing a one-dimensional magnitude limit or a hard completeness cut, \cfour~models the probability that each synthetic cluster would be recovered by the observational LEGUS pipeline. For a synthetic library cluster $j$ with photometric vector $\boldsymbol{m}_j$ and a filter set $\mathcal{F}$, the network computes $\widehat{P}_{\rm obs,\mathcal{F}}(\boldsymbol{m}_j)$, which in turn
enters directly into the library weights in \autoref{eq:library_weights}.

The neural network in \cfour~requires training data, which we generate for each field within each galaxy using field-specific artificial star test realisations following the procedure described in \citet{Tang2026}. Synthetic clusters are drawn randomly from the \slug~library, which provides their physical properties and multi-band photometry, and are injected into the real observed \textit{HST} images. The training-set size 20,000--50,000 clusters per field, which we set based on the convergence experiments in \citet{Tang2026}, which show that stable completeness estimates are reached with a few $10^4$ injected clusters per field. We process each artificial training cluster through the same detection, photometry, matching, and catalogue-selection workflow as the real cluster candidates to see if it is recovered, and then train our neural networks (one per LEGUS field, and thus one per filter set) to learn detection probability as a function of photometric magnitude from these experiments.

We do not retrain \cfour~for radial subdivisions of a galaxy. Because the artificial clusters are injected into each field at positions drawn from the observed galaxy light distribution, the resulting completeness model samples, and thus marginalises over, the range of crowding and galaxy-background conditions within each galaxy. We therefore expect the galactic-level completeness model to provide an adequate approximation for the radial subsamples. Dividing the existing artificial-cluster sample among smaller subregions would increase finite-sample noise in the inferred completeness, whereas retaining the same training-set size in every subregion would substantially increase the computational cost.

\subsection{Cluster demographic models}

The remaining step in specifying our model is to choose functional forms for $f(M, T, A_V \mid \boldsymbol{\theta})$, the function that describes the intrinsic distribution of cluster properties, whose parameters $\boldsymbol{\theta}$ we wish to infer. We consider two model families: mass-independent disruption (MID) and mass-dependent disruption (MDD), defined by \citet{Krumholz2019SLUGIV}. 

\subsubsection{Mass-independent disruption model (MID)} 
In the MID case, the present-day joint mass--age distribution is
\begin{equation}
    \frac{{\rm d}^2N}{{\rm d}M\,{\rm d}T} \propto
    M^{\alpha_M}\exp\left(-\frac{M}{M_{\rm break}}\right)
    \max(T,T_{\rm MID})^{\alpha_T}.
    \label{eq:mid_model}
\end{equation}
In this family of model, we have four parameters to infer: $M_\mathrm{break}$, $\alpha_M$, $T_\mathrm{MID}$, and $\alpha_T$. These describe, respectively, the slope of the powerlaw portion of the cluster mass function, the mass at which this powerlaw behaviour gives way to an exponential truncation, the age below which clusters do not disrupt (and thus the cluster age distribution is flat), and the slope of the age distribution for clusters older than the age at which disruption begins. In addition to these fundamental quantities of our fit, we also report the effective cluster mass function slope evaluated at $10^4\,M_\odot$ for direct comparison with previous work; this quantity is
\begin{equation}
    \alpha_{M_4} = \alpha_M - \frac{10^4\,\mathrm{M}_\odot}{M_\mathrm{break}}.
\end{equation}

\subsubsection{Mass-Dependent Disruption model (MDD)}
For the MDD model, we adopt a mass-loss prescription in which the disruption rate depends on the current cluster mass \citep{2007MDD,2021MDD},
\begin{equation}
    \frac{\mathrm{d}M}{\mathrm{d}T} \propto -M^{\gamma_{\mathrm{MDD}}},
\end{equation}
so a cluster with initial mass $M_i$ and age $T$ has present-day mass
\begin{equation}
    \label{eq:mdd_mass_loss}
    M = M_i
    \left[
    1 - \gamma_{\mathrm{MDD}}
    \left(\frac{M_0}{M_i}\right)^{\gamma_{\mathrm{MDD}}}
    \frac{T}{T_{\mathrm{MDD},0}}
    \right]^{1/\gamma_{\mathrm{MDD}}},
\end{equation}
where $T_{\mathrm{MDD},0}$ is the disruption time of a cluster with reference mass $M_0$. Combining this evolution with the Schechter initial mass function, the present-day mass--age distribution becomes
\begin{equation}
    \label{eq:mass_age_mdd}
    \frac{\mathrm{d}^{2}N}{\mathrm{d}M\,\mathrm{d}T}
    \propto
    M^{\alpha_M}
    \eta^{\alpha_M + 1 - \gamma_{\mathrm{MDD}}}
    \exp\left(-\eta\frac{M}{M_{\mathrm{break}}}\right),
\end{equation}
where
\begin{equation}
    \eta(M,T) =
    \left[
    1 + \gamma_{\mathrm{MDD}}
    \left(\frac{M_0}{M}\right)^{\gamma_{\mathrm{MDD}}}
    \frac{T}{T_{\mathrm{MDD},0}}
    \right]^{1/\gamma_{\mathrm{MDD}}}
\end{equation}
is the ratio between the initial and present-day cluster mass.

The MDD model has four demographic parameters to infer: $\alpha_M$ and $ M_{\mathrm{break}}$, which have the same meaning as for the MID model, $\gamma_{\mathrm{MDD}}$, which sets the mass dependence of the disruption rate, and $\log(T_{\mathrm{MDD},0}/\mathrm{yr})$, the disruption timescale at the reference mass. We fix $M_0 = 100\,M_\odot$ without loss of generality, since the distribution depends only on the combination $M_0^{\gamma_{\mathrm{MDD}}}/T_{\mathrm{MDD},0}$, not on $M_0$ alone. For comparison with previous work, we also report the disruption time for a $10^4\,M_\odot$ cluster,
\begin{equation}
    t_4 = 10^{2\gamma_{\mathrm{MDD}}} T_{\mathrm{MDD},0}.
\end{equation}

\subsubsection{Dust extinction}

In addition to the models for the mass and age distribution, $f$ also includes the PDF of extinctions. Because the intrinsic extinction distribution is not known, we treat this PDF $p(A_V)$ as a nuisance parameter of our model. For both disruption models, we parameterise $p(A_V)$ as a piecewise-linear function over $0 \leq A_V \leq 3$ mag, with the piecewise segments anchored at values of $A_V$ separated by $\Delta A_V = 0.5$ mag whose values are free parameters in $\boldsymbol{\theta}$ -- see \citet{Krumholz2019SLUGIV} for a detailed description. Combining mass, age, and extinction, the final joint function to be fit is
\begin{equation}
    f(M,T,A_V \mid \boldsymbol{\theta})
    \propto
    \frac{\mathrm{d}^{2}N}{\mathrm{d}M\,\mathrm{d}T}
    p(A_V),
    \label{eq:joint_mass_age_av}
\end{equation}
where $\mathrm{d}^{2}N/\mathrm{d}M\,\mathrm{d}T$ is given by the corresponding mass--age distribution for the MID or MDD model.


\subsection{Calculation of posteriors}
\label{ssec:posteriors}

We use 
the affine-invariant ensemble sampler \texttt{emcee} \citep{emcee} to sample the posterior distribution given by \autoref{eq:posterior} for each of our target galaxies, combining all the LEGUS fields in that galaxy into a single catalogue for those galaxies covered by more than one LEGUS field (\autoref{tab:galaxy_properties}). We also carry out separate \texttt{emcee} runs for some galactic sub-regions, as discussed further in \autoref{ssec:env_dependence}. In all cases we run separate chains for both the MID and MDD model.

For these runs we adopt broad, flat priors for all parameters. Specifically, for both model families, we adopt flat priors in the CMF slope over the range $-3 < \alpha_M < 0$, and for the log of the CMF break mass over $2 < \log_{10}(M_{\rm break}/M_\odot) < 8$. For the MID model, we adopt flat priors over $-3 < \alpha_T < 0$ and $5 < \log_{10}(T_{\rm MID}/{\rm yr}) < 10$, and for the MDD model we adopt flat priors over $0 < \gamma_{\rm MDD} < 1$ and $4 < \log_{10}(T_{{\rm MDD},0}/{\rm yr}) < 10$.

We initialise each chain with 100 walkers and run for 2000 iterations. The walkers are initialised in broad regions within the prior volume. We discard the first 500 iterations as burn-in and construct the reported marginal posterior intervals from the remaining samples. We assess convergence by checking that the parameter chains show no remaining long-term drift, visually inspecting the walker traces, and calculating the integrated autocorrelation time, $\tau_{\rm int}$, for each parameter. We require $N_{\rm post} \geq 50\,\tau_{\rm int}$ for each parameter, where $N_{\rm post}$ is the number of post-burn-in samples; our 1500 post-burn-in iterations is in all cases sufficient to satisfy this requirement.

Once we have obtained posterior samples directly for both the MID and MDD models for a given cluster catalogue, we compare the two models using the Akaike information criterion,
\begin{equation}
    {\rm AIC} = 2k - 2\ln \hat{\mathcal{L}},
\end{equation}
where $k$ is the number of free parameters of the fit (which is the same for both models, and thus does not influence the model comparison) and $\hat{\mathcal{L}}$ is the maximum likelihood reached by the chain. We convert the AIC values into relative Akaike weights,
\begin{equation}
    w_i =
    \frac{\exp(-\Delta_i/2)}
    {\sum_j \exp(-\Delta_j/2)},
\end{equation}
where $\Delta_i = {\rm AIC}_i - \min_j({\rm AIC}_j)$, and $i$ and $j$ run over the models MID and MDD. The Akaike weight $w_i$ can be interpreted as the relative support for model $i$ among the two models; values close to unity indicate strong preference for that model, while values close to $0.5$ indicate that the data do not clearly distinguish between MID and MDD.

\section{Results}
\label{sec:results}

We first in \autoref{sec:demographic_results} examine results of the full-galaxy catalogues, and then in \autoref{ssec:env_dependence} we present results for subregions within selected individual galaxies.

\subsection{Cluster demographics: the galactic scale}
\label{sec:demographic_results}

\autoref{Tab:bestfit_params} reports the posterior medians and $16^{\rm th}$--$84^{\rm th}$ percentile intervals for the best-fitting model for each galaxy in our sample. Since several parameters exhibit substantial covariance, particularly the CMF slope and $M_{\rm break}$, the marginal intervals in \autoref{Tab:bestfit_params} should be interpreted together with the corresponding joint posterior distributions. The full posterior distributions, parameter constraints for both the MID and MDD models, and the associated model-comparison statistics are provided in the online supplemental material, but we discuss illustrative examples for select galaxies below.

\begin{figure}
    \centering
    \includegraphics[width=0.5\textwidth]{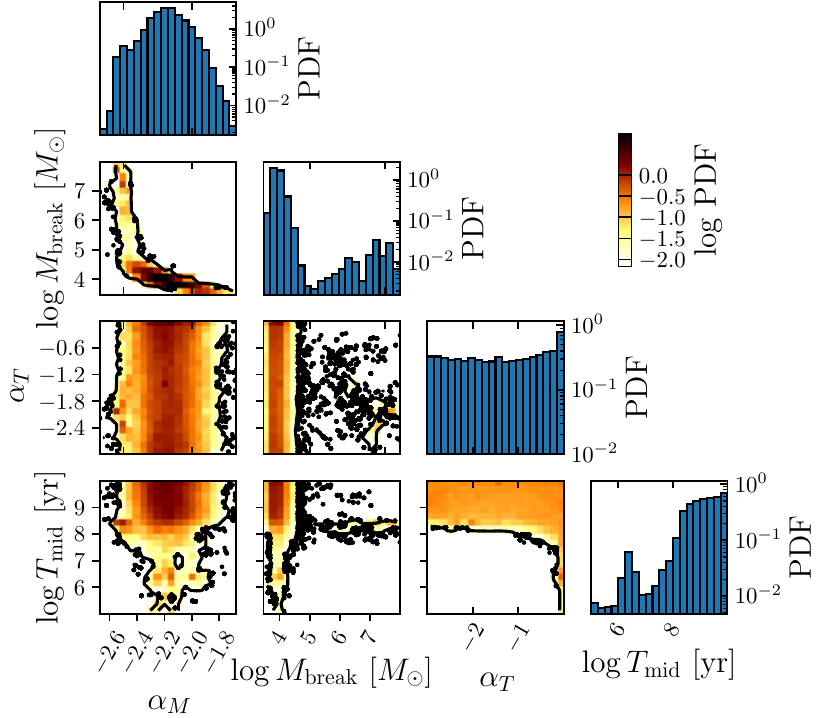}
    \caption{Corner plot showing the posterior probability distributions for the inferred demographic parameters of the NGC~3344 cluster population. The diagonal panels show the 1D marginal posterior distributions, while the off-diagonal panels show the corresponding 2D posterior probability densities. The colour scale of 2D PDFs represents the logarithmic values of the posterior probability density. The parameters shown are the cluster mass function slope, $\alpha_M$, the characteristic truncation mass, $\log M_{\rm break}$, the age function slope, $\alpha_T$, and the minimum age of the power law regime, $\log T_{\rm MID}$. The outermost contour encloses 99 percent of the posterior samples, and black points indicate samples lying outside this contour. The axis limits are rounded to the prior ranges for clarity of presentation. The strong, localized peak in the $M_{\rm break}$ posterior indicates a well-constrained finite truncation mass.}
    \label{fig:NGC3344}
\end{figure}

\begin{figure}
    \centering
    \includegraphics[width=0.5\textwidth]{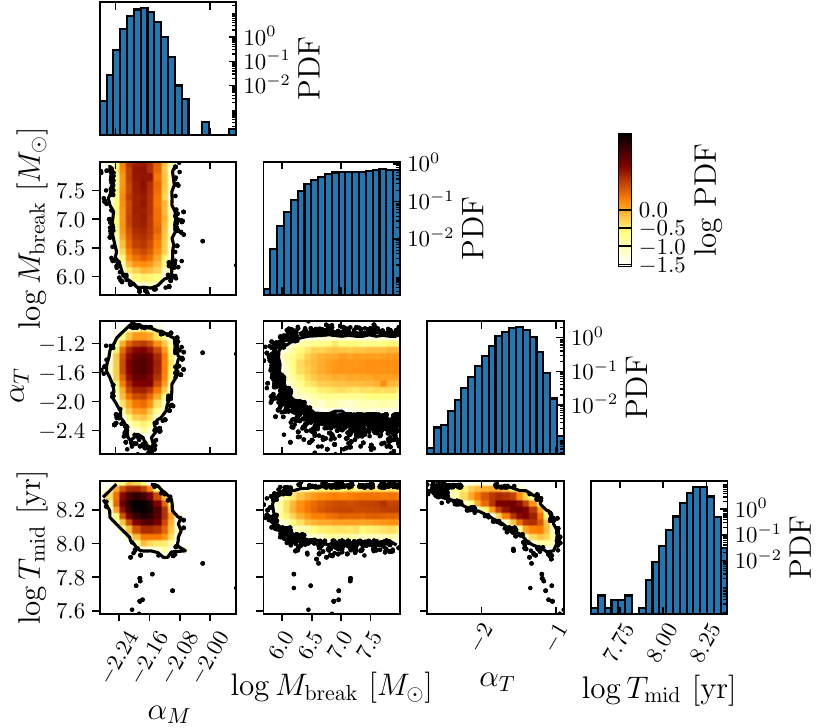}
    \caption{Same as \autoref{fig:NGC3344}, except for NGC~628. Its cluster mass function resembles a pure power law, as indicated by the posterior extending towards high values of $M_{\rm break}$.}
    \label{fig:NGC628}
\end{figure}

The inferred cluster mass functions are generally consistent with an approximately $M^{-2}$ form. For the full-galaxy catalogues, the asymptotic low-mass slope spans approximately $-2.24 \lesssim \alpha_M \lesssim -1.46$, with most systems lying close to $\alpha_M=-2$. However, $\alpha_M$ and the characteristic break mass $M_{\rm break}$ are strongly covariant: a shallower asymptotic slope combined with a lower $M_{\rm break}$ can produce nearly the same CMF over the mass range sampled by the observations as a steeper slope with a higher break mass. Consequently, the effective logarithmic slope evaluated at $10^4\,M_\odot$, $\alpha_{M4}$, is often more tightly constrained than either $\alpha_M$ or $M_{\rm break}$ separately. Several catalogues show posterior support concentrated at a finite characteristic mass, including NGC~3344 (see \autoref{fig:NGC3344}), NGC~3627, NGC~5194--NGC~5195, the combined NGC~5457 catalogue, and NGC~6503. By contrast, NGC~628 (see \autoref{fig:NGC628}), NGC~3738, NGC~4449, NGC~7793, and several of the resolved pointings exhibit broad, multimodal, or prior-limited $M_{\rm break}$ posteriors that extend to masses well above the observed cluster population. These cases remain consistent with an effectively untruncated power law. 
As we will show in \autoref{ssec:env_dependence}, some galaxies show significant sub-galactic-scale variation in their CMFs, and thus
a single full-galaxy CMF may in some cases represent an effective description of a mixture of regionally distinct cluster populations.

The model comparison shows strong evidence for mass-independent disruption in nine of the twelve galaxy-scale catalogues. The MID model has an Akaike weight $w_{\rm MID}> 0.9$ for NGC~1313, NGC~3627, NGC~4449, NGC~5194--NGC~5195, NGC~5253, NGC~5457, NGC~628, NGC~6503 and NGC~7793. For NGC~1566 and NGC~3738, MID provides a marginally better fit whereas MDD is marginally favoured for NGC~3344. However, all three galaxies have $w_{\rm MID}\simeq w_{\rm MDD}$ and $\log_{10}(w_{\rm MID}/w_{\rm MDD})\approx0$, indicating that the two prescriptions provide statistically indistinguishable fits to the current data. This does not imply that the underlying disruption mechanisms are physically equivalent, but that the observed photometry provides insufficient leverage to favour one parametrization over the other. We can understand the reason that the data cannot distinguish between the MID and MDD disruption mechanisms in these cases by noting that both the MID and MDD fits recover nearly identical CMF parameters, while the parameters describing the disruption timescale -- $T_\mathrm{MID}$ for the MID model and $t_4$ for the MDD model -- both have 50th percentile values well above 1 Gyr in these three galaxies. Thus the conclusion to which we are led by the data is that in these three systems there is little evidence for cluster disruption \textit{at all}, at least over the timescales to which our data are sensitive -- and in the absence of any measurement of cluster disruption, we obviously cannot say whether the disruption is mass-dependent or mass-independent.

Within the MID parametrization, the cluster age function is determined jointly by the age at which disruption begins to affect the observed distribution, $T_{\rm MID}$, and the subsequent CAF slope, $\alpha_T$. Values $-1<\alpha_T<0$ correspond to a relatively shallow decline, whereas $\alpha_T\lesssim-1$ indicates a steeper decline. However, $\alpha_T$ and $T_{\rm MID}$ are strongly covariant: when $T_{\rm MID}$ approaches or exceeds the oldest ages effectively sampled by a catalogue, the fitted value of $\alpha_T$ has little observable effect and this leads to very broad posteriors. NGC~1313 provides the clearest example of a steeply declining CAF, with disruption beginning at $T_{\rm MID}\simeq250$~Myr and $\alpha_T\simeq-2.5$. NGC~628 similarly favours disruption beginning at $T_{\rm MID}\simeq160$~Myr, while NGC~3627 favours $T_{\rm MID}\simeq680$~Myr, both followed by a moderately steep decline with $\alpha_T\simeq-1.5$. By contrast, NGC~5253 favours disruption at approximately $1$~Myr with a shallower subsequent slope, $\alpha_T\simeq-0.7$. NGC~4449, NGC~6503, and NGC~7793 favour earlier disruption, from $\sim1$ to a few tens of Myr, with shallower declines with $\alpha_T\simeq-0.5$ to $-0.7$. NGC~5194--NGC~5195 has a broad or multimodal $T_{\rm MID}$ posterior and therefore permits both early- and late-disruption solutions; its nominally steep value of $\alpha_T\simeq-1.6$ is correspondingly less secure. For NGC~1566, NGC~3344, NGC~3738, and NGC~5457, the posterior places disruption at $\gtrsim1$~Gyr and remains broad in both parameters, so the CAF is consistent with being approximately flat over most of the observed age range and provides no evidence for rapid early disruption.

\begin{table*}
\centering
\caption{Posterior constraints for the AIC-favoured population model of each galaxy-scale catalogue. For each galaxy, the model with the larger Akaike weight is listed. Posterior quantities are reported as ${q_{50}}^{+(q_{84}-q_{50})}_{-(q_{50}-q_{16})}$, where $q_N$ denotes the $N$th percentile of the marginalized posterior distribution. The quoted ranges therefore correspond to 68 percent confidence intervals. A dash indicates that a parameter does not apply to the selected model. The complete results for both models and all sub-galaxy catalogues are given in online-only supplementary materials. The quantity $\log_{10}(w_{\rm MID}/w_{\rm MDD})$ gives the relative Akaike weight comparison for the mass independent disruption (MID) and the mass dependent disruption (MDD) models; positive values favour MID, whereas negative values favour MDD. Values near zero indicate that MID and MDD have comparable fits to data.}
\label{Tab:bestfit_params}
\scriptsize
\setlength{\tabcolsep}{2pt}
\renewcommand{\arraystretch}{1.2}
\begin{tabular}{llccccccccc}
\hline
Catalogue & Model
& $\log_{10}(w_{\rm MID}/w_{\rm MDD})$
& $\alpha_M$
& $\log_{10}(M_{\rm break}/M_\odot)$
& $\alpha_{M_4}$
& $\alpha_T$
& $\log_{10}(T_{\rm MID}/{\rm yr})$
& $\log_{10}(T_{\rm MDD,0}/{\rm yr})$
& $\gamma_{\rm MDD}$
& $t_4$ (Myr) \\
\hline
NGC~0628
& MID
& $8.77$
& $-2.18^{+0.02}_{-0.02}$
& $7.25^{+0.51}_{-0.60}$
& $-2.18^{+0.02}_{-0.02}$
& $-1.54^{+0.18}_{-0.21}$
& $8.20^{+0.05}_{-0.06}$
& --
& --
& -- \\

NGC~1313
& MID
& $7.87$
& $-1.93^{+0.05}_{-0.05}$
& $5.11^{+0.19}_{-0.16}$
& $-2.01^{+0.08}_{-0.08}$
& $-2.52^{+0.43}_{-0.33}$
& $8.39^{+0.03}_{-0.04}$
& --
& --
& -- \\

NGC~1566
& MID
& $0.07$
& $-1.85^{+0.03}_{-0.03}$
& $4.82^{+0.04}_{-0.04}$
& $-2.00^{+0.04}_{-0.04}$
& $-1.39^{+1.05}_{-1.12}$
& $9.42^{+0.40}_{-0.39}$
& --
& --
& -- \\

NGC~3344
& MDD
& $-0.03$
& $-2.18^{+0.12}_{-0.12}$
& $3.94^{+0.20}_{-0.14}$
& --
& --
& --
& $9.19^{+0.58}_{-0.85}$
& $0.69^{+0.23}_{-0.36}$
& $16328^{+72163}_{-12429}$ \\

NGC~3627
& MID
& $2.86$
& $-1.46^{+0.05}_{-0.05}$
& $4.90^{+0.05}_{-0.04}$
& $-1.59^{+0.06}_{-0.06}$
& $-1.65^{+0.85}_{-0.85}$
& $8.83^{+0.13}_{-0.11}$
& --
& --
& -- \\

NGC~3738
& MID
& $0.05$
& $-2.24^{+0.08}_{-0.06}$
& $5.82^{+1.54}_{-0.79}$
& $-2.26^{+0.09}_{-0.14}$
& $-1.34^{+1.05}_{-1.22}$
& $9.12^{+0.62}_{-0.70}$
& --
& --
& -- \\

NGC~4449
& MID
& $1.03$
& $-1.85^{+0.06}_{-0.06}$
& $5.77^{+0.44}_{-0.22}$
& $-1.87^{+0.07}_{-0.07}$
& $-0.64^{+0.04}_{-0.05}$
& $6.39^{+0.05}_{-0.05}$
& --
& --
& -- \\

NGC~5194--NGC~5195
& MID
& $4.37$
& $-2.05^{+0.08}_{-0.04}$
& $4.74^{+0.09}_{-0.10}$
& $-2.23^{+0.11}_{-0.09}$
& $-1.58^{+1.49}_{-0.95}$
& $8.57^{+0.10}_{-2.17}$
& --
& --
& -- \\

NGC~5253
& MID
& $3.73$
& $-1.61^{+0.15}_{-0.16}$
& $5.93^{+1.24}_{-0.42}$
& $-1.62^{+0.16}_{-0.18}$
& $-0.69^{+0.12}_{-0.14}$
& $5.90^{+0.63}_{-0.68}$
& --
& --
& -- \\

NGC~5457
& MID
& $2.40$
& $-2.04^{+0.05}_{-0.05}$
& $4.61^{+0.23}_{-0.08}$
& $-2.29^{+0.15}_{-0.10}$
& $-0.75^{+0.43}_{-1.54}$
& $9.24^{+0.46}_{-0.76}$
& --
& --
& -- \\

NGC~6503
& MID
& $5.32$
& $-1.85^{+0.15}_{-0.12}$
& $4.34^{+0.31}_{-0.42}$
& $-2.31^{+0.38}_{-0.87}$
& $-0.48^{+0.21}_{-0.22}$
& $7.50^{+1.19}_{-0.13}$
& --
& --
& -- \\

NGC~7793
& MID
& $1.38$
& $-2.24^{+0.32}_{-0.21}$
& $4.38^{+3.48}_{-0.46}$
& $-2.66^{+0.74}_{-1.00}$
& $-0.61^{+0.47}_{-1.39}$
& $7.57^{+1.64}_{-0.91}$
& --
& --
& -- \\
\hline
\end{tabular}
\end{table*}

\subsection{Cluster demographics: the sub-galactic scale}
\label{ssec:env_dependence}

\begin{table*}
\centering
\caption{MID cluster-mass-function parameters for the radial subdivisions
defined by $\beta=0.3$ (see \autoref{ssec:env_dependence}) and for the observational-field subdivisions of NGC~628, NGC~5457 and NGC~5194--NGC~5195. The reported values are the medians and $16^{\rm th}$--$84^{\rm th}$ percentile intervals of the marginalized posteriors. The column $N_{\rm cl}$ gives the number of clusters included in each region.
\label{Tab:regional_cmf_params}
}
\scriptsize
\setlength{\tabcolsep}{5pt}
\renewcommand{\arraystretch}{1.2}
\begin{tabular}{lllccc}
\hline\hline
Galaxy & Division & Region/field & $N_{\rm cl}$ & $\alpha_M$
& $\log_{10}(M_{\rm break}/M_\odot)$ \\
\hline
NGC~0628 & $\beta=0.3$
& $R_{\rm gal}<1.6\,\mathrm{kpc}$
& 145 & $-2.32^{+0.13}_{-0.16}$ & $7.20^{+0.55}_{-0.65}$ \\
& & $R_{\rm gal}\geq1.6\,\mathrm{kpc}$
& 1093 & $-2.14^{+0.03}_{-0.03}$ & $6.97^{+0.71}_{-0.83}$ \\
\hline
& Observational fields & C
& 990 & $-2.19^{+0.03}_{-0.03}$ & $7.35^{+0.44}_{-0.51}$ \\
& & E
& 249 & $-1.62^{+0.11}_{-0.12}$ & $4.23^{+0.12}_{-0.09}$ \\
\hline\hline
NGC~5194--NGC~5195 & $\beta=0.3$
& $R_{\rm gal}<1.6\,\mathrm{kpc}$
& 86 & $-1.68^{+0.14}_{-0.12}$ & $5.42^{+1.04}_{-0.35}$ \\
& & $R_{\rm gal}\geq1.6\,\mathrm{kpc}$
& 1142 & $-2.04^{+0.04}_{-0.04}$ & $4.70^{+0.08}_{-0.08}$ \\
\hline\hline
NGC~5457 & $\beta=0.3$
& $R_{\rm gal}<2.4\,\mathrm{kpc}$
& 273 & $-2.41^{+0.05}_{-0.04}$ & $7.08^{+0.63}_{-0.62}$ \\
& & $R_{\rm gal}\geq2.4\,\mathrm{kpc}$
& 871 & $-1.83^{+0.05}_{-0.05}$ & $4.37^{+0.06}_{-0.05}$ \\
\hline
& Observational fields & C
& 413 & $-2.28^{+0.04}_{-0.04}$ & $6.80^{+0.82}_{-0.62}$ \\
& & SE
& 442 & $-1.80^{+0.08}_{-0.08}$ & $4.29^{+0.10}_{-0.08}$ \\
& & NW1
& 217 & $-1.86^{+0.10}_{-0.10}$ & $4.47^{+0.14}_{-0.10}$ \\
& & NW2
& 56 & $-2.28^{+0.17}_{-0.13}$ & $5.02^{+1.86}_{-0.45}$ \\
& & NW3
& 17 & $-1.86^{+0.98}_{-0.45}$ & $4.03^{+2.58}_{-0.64}$ \\
\hline\hline
\end{tabular}
\end{table*}

We next test for sub-galactic variations in cluster demographics. Since we are interested in the possible effects of sub-galactic scale processes such as shear, we focus on the three LEGUS galaxies for which suitable rotation curves are readily available in the literature: NGC~5457, NGC~5194--NGC~5195, and NGC~628 \citep{SuwannajakTanLeroy2014}. The definition of the $\beta$-based division and the calculation of the corresponding dividing radii are described in \autoref{sec:environmental_measurements}.

\begin{figure*}
    \centering
    \includegraphics[width=\textwidth]{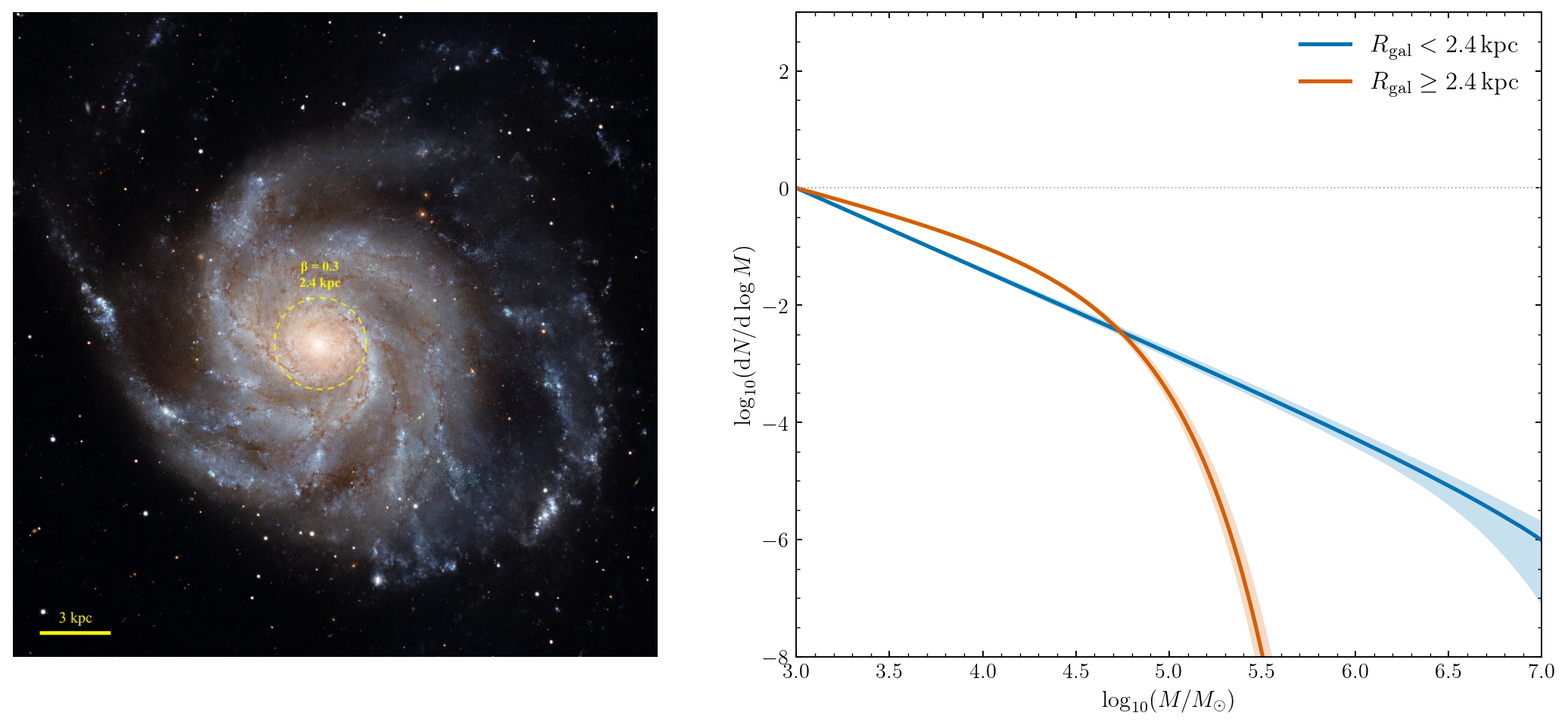}
    \caption{Sub-galactic CMF comparison for NGC~5457. Left: a false colour HST image of NGC~5457 (source: ESA) with a dashed yellow circle indicating the boundary between our inner and outer galaxy test regions; the circle lies at $R_{\rm gal}=2.4$~kpc, and corresponds to the galactocentric radius at which the rotation curve index $\beta=0.3$. Right: posterior CMFs for the inner and outer regions, normalized to unity at $10^3\,M_\odot$. Solid curves denote medians, and shaded regions enclose the $16^{\rm th}$ - $84^{\rm th}$ percentile intervals obtained by Monte Carlo sampling of the joint posterior distributions of $\alpha_M$ and $M_{\rm break}$.}
    \label{fig:NGC5457_beta}
\end{figure*}

Because the LEGUS observations of NGC 628 and NGC~5457 cover multiple pointings (see \autoref{tab:galaxy_properties}), as a further test we also carry out separate fits for each pointing. This experiment is particularly powerful in NGC~5457, which is covered by five pointings with comparable projected areas that sample distinct radial environments, providing a finer-grained test of the variation than the two-region split. We show the fields covered by each of our tests in the left panels of \autoref{fig:NGC5457_beta} - \autoref{fig:NGC628_beta}.

We report best-fit MID cluster mass function (CMF) parameters for each of our sub-galactic test fields in \autoref{Tab:regional_cmf_params}; because MID is the preferred prescription for all three galaxies, we omit tests against MDD models here. The right panels of \autoref{fig:NGC5457_beta} - \autoref{fig:NGC628_beta} show the posterior CMFs for each of the test regions, normalised to the same amplitude at $10^3\,M_\odot$ for convenience of comparison. Solid curves denote the posterior medians, and the shaded regions enclose the 16th--84th percentile intervals obtained by Monte Carlo sampling of the joint posterior distributions of $\alpha_M$ and $M_{\rm break}$. We next discuss each galaxy individually.

\subsubsection{NGC~5457}
\begin{figure*}
    \centering
    \includegraphics[width=\textwidth]{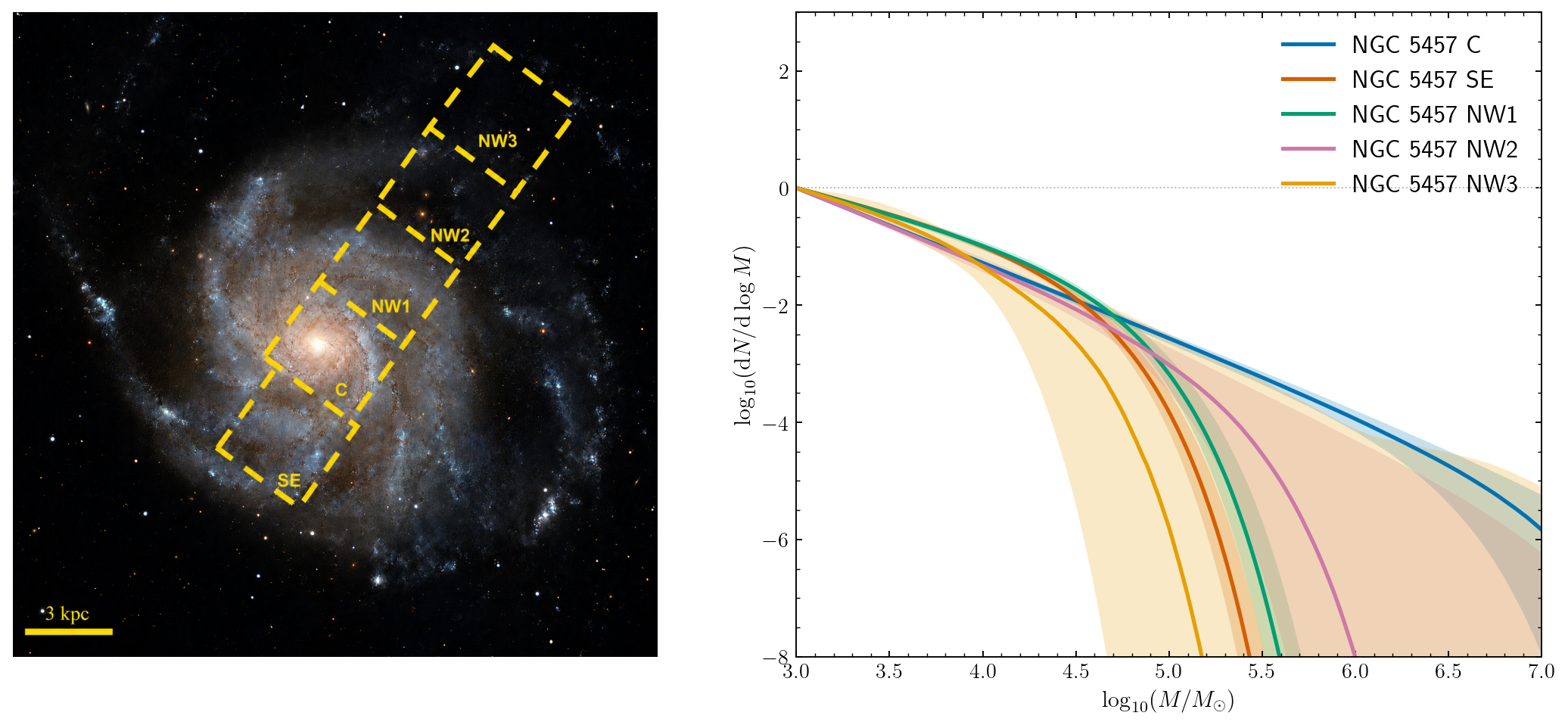}
    \caption{Same as \autoref{fig:NGC5457_beta}, but for the five LEGUS pointings in NGC~5457. In the left panel, the yellow dashed boxes mark the central (C), south-eastern (SE), and three north-western (NW1--NW3) fields on an HST false-colour image; the right panel shows their corresponding posterior CMFs.}
    \label{fig:NGC5457_fields}
\end{figure*}

NGC~5457 shows the strongest radial change in CMF shape. The inner region has $\alpha_M=-2.41^{+0.05}_{-0.04}$ and a high truncation scale, $\log_{10}(M_{\rm break}/M_\odot)=7.08^{+0.63}_{-0.62}$, placing the cutoff above the observed mass range and making the CMF effectively a pure power law but with a somewhat steeper slope. The outer region instead has a shallower slope, $\alpha_M=-1.83\pm0.05$, followed by a well-constrained truncation at $\log_{10}(M_{\rm break}/M_\odot)=4.37^{+0.06}_{-0.05}$ (\autoref{fig:NGC5457_beta}).

The five-pointing comparison in \autoref{fig:NGC5457_fields} shows that the variation is not perfectly monotonic with position. The central field favours a steep CMF with a high truncation scale, $\log_{10}(M_{\rm break}/M_\odot)=6.80^{+0.82}_{-0.62}$, consistent with the inner region in our radial division -- not a surprising result given the large overlap between the inner and C regions. By contrast, the SE and NW1 fields have lower, well-constrained values of $4.29^{+0.10}_{-0.08}$ and $4.47^{+0.14}_{-0.10}$, respectively. NW2 and NW3 have $\log_{10}(M_{\rm break}/M_\odot)$ values are essentially unconstrained, $5.02^{+1.86}_{-0.45}$ and $4.03^{+2.58}_{-0.64}$ respectively, which is not surprising given that they contain very few clusters. Nonetheless, the differences between C, SE, and NW1, as well as between inner and outer, all of which are well-constrained, provide clear evidence of an environmental imprint on cluster formation. The robust result is a suppression of the high-mass CMF in fields away from the galactic centre. 

\subsubsection{NGC~5194--NGC~5195}
\begin{figure*}
    \centering
    \includegraphics[width=\textwidth]{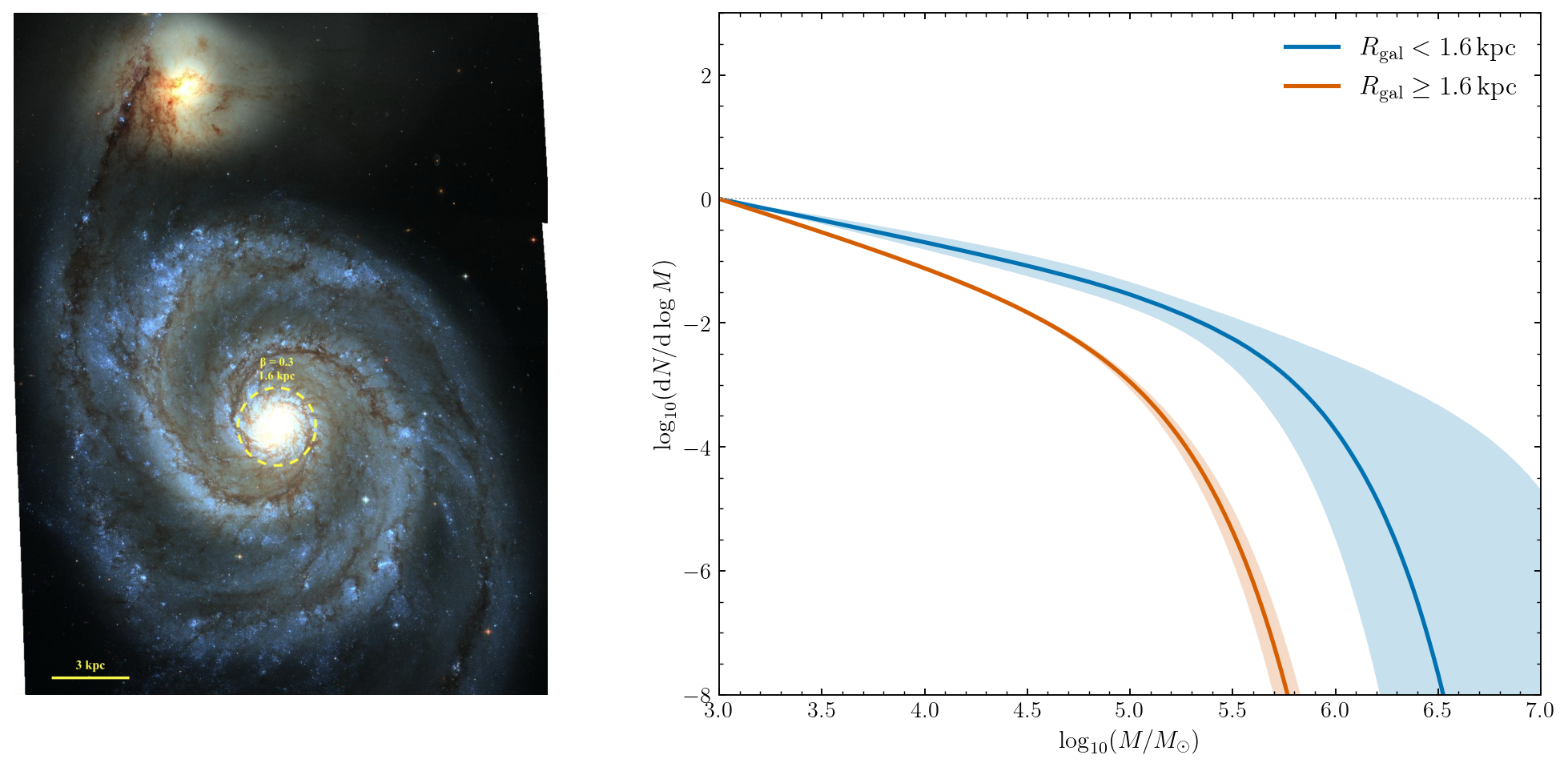}
    \caption{Same as \autoref{fig:NGC5457_beta}, but for NGC~5194--NGC~5195.}
    \label{fig:NGC5194_5195_beta}
\end{figure*}

NGC~5194--NGC~5195 also shows a clear radial variation. In \autoref{fig:NGC5194_5195_beta}, the inner $R_{\rm gal}<1.6$~kpc CMF has a shallow slope, $\alpha_M=-1.68^{+0.14}_{-0.12}$, and a comparatively high but only weakly-constrained truncation scale, $\log_{10}(M_{\rm break}/M_\odot)=5.42^{+1.04}_{-0.35}$. The outer CMF has a slightly steeper slope, $\alpha_M=-2.04\pm0.04$, and a much lower, tightly constrained truncation mass $\log_{10}(M_{\rm break}/M_\odot)=4.70\pm0.08$. Thus both fitted parameters act in the same direction: relative to the outer disc, the inner population is richer in high-mass clusters and its cutoff occurs at a higher mass.

\subsubsection{NGC~628}
\begin{figure*}
    \centering
    \includegraphics[width=\textwidth]{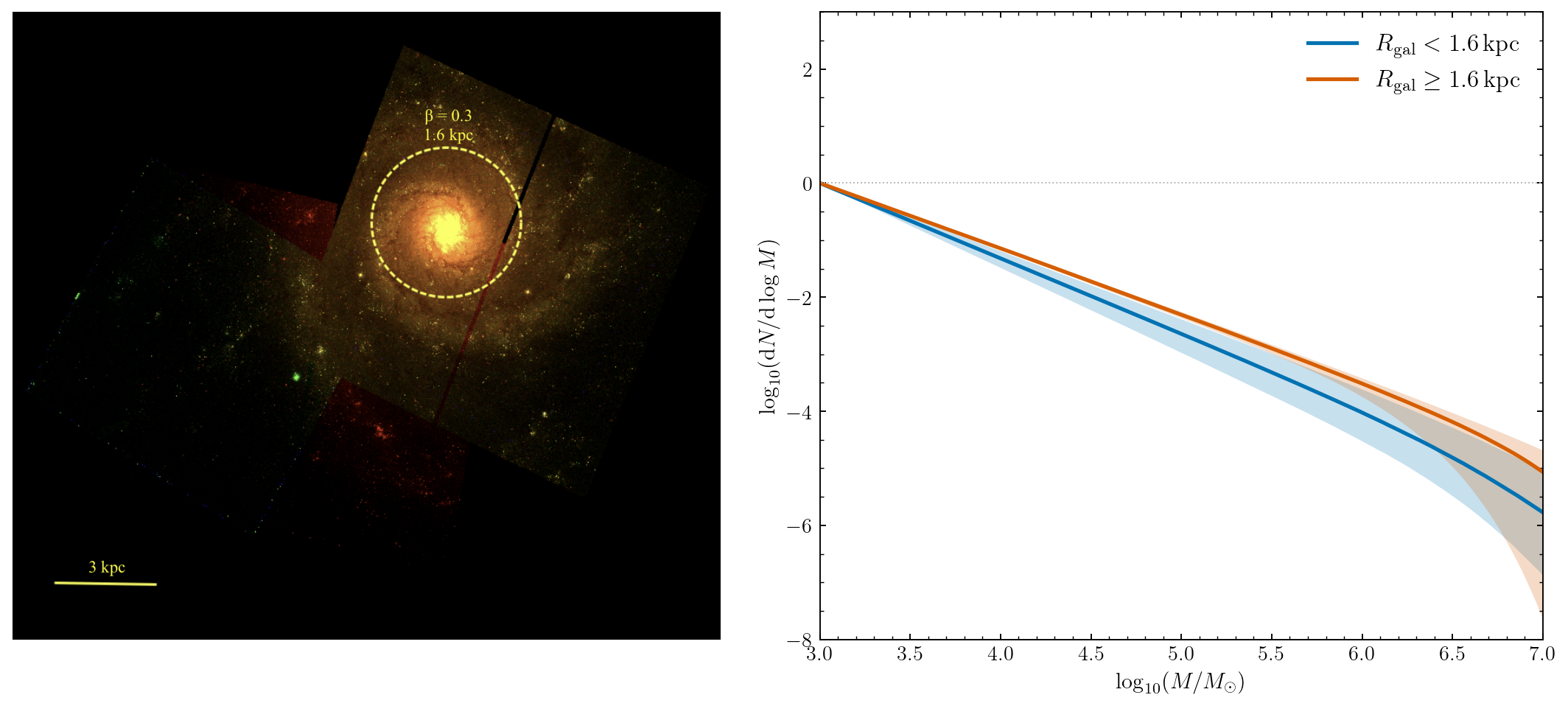}
    \caption{Same as \autoref{fig:NGC5457_beta}, but for NGC~628.}
    \label{fig:NGC628_beta}
\end{figure*}

NGC~628 shows very different radial dependence than NGC 5457 or 5194-5195. As shown in \autoref{fig:NGC628_beta}, its inner and outer radial CMFs are similar: the inferred slopes are $-2.32^{+0.13}_{-0.16}$ and $-2.14\pm0.03$, respectively, while both regions favour high and mutually consistent truncation scales, with $\log_{10}(M_{\rm break}/M_\odot)=7.20^{+0.55}_{-0.65}$ inside and $6.97^{+0.71}_{-0.83}$ outside. The outer curve is therefore not truncated at low mass. By contrast, there is a clear difference between the two NGC~628 fields:
the central field is consistent with a steep, effectively untruncated CMF, whereas the eastern field favours a shallow slope and a low $M_{\rm break}$. Compared with \citet{Tang2024} on NGC~628 using the same cluster catalogue, the current analysis retains the steep CMF slope of $\sim-2.2$ but favours an effectively untruncated CMF for the full and central samples, removing the previous degeneracy with a low-$M_{\rm break}$ solution near $10^{4.5}\,M_\odot$. This difference likely arises from the \cfour~completeness treatment, which allows approximately 100 additional clusters previously excluded because of their low estimated completeness to be incorporated. Despite the enlarged sample, the AIC decreases from $5916$ to $5520$, supporting the conclusion that the revised completeness model provides a better fit to the observations.

\section{Discussion}
\label{sec:discussion}

Here we discuss some of the implications of our results, with particular attention to how they compare to earlier work on star cluster demographics and their dependence on environment.

The key result of this work is not simply that some galaxies show truncated mass functions and others do not; it is that a uniform forward-modelling treatment reveals substantially more uncertainty in $M_{\rm break}$ than is often apparent from previous analyses, and that this uncertainty is itself environmentally dependent.

\subsection{\texorpdfstring{The $\Sigma_{\rm SFR}$--$M_{\rm break}$ relation}{Sigma SFR--M break relation}}

\begin{figure*}
    \centering
    \includegraphics[width=\textwidth]{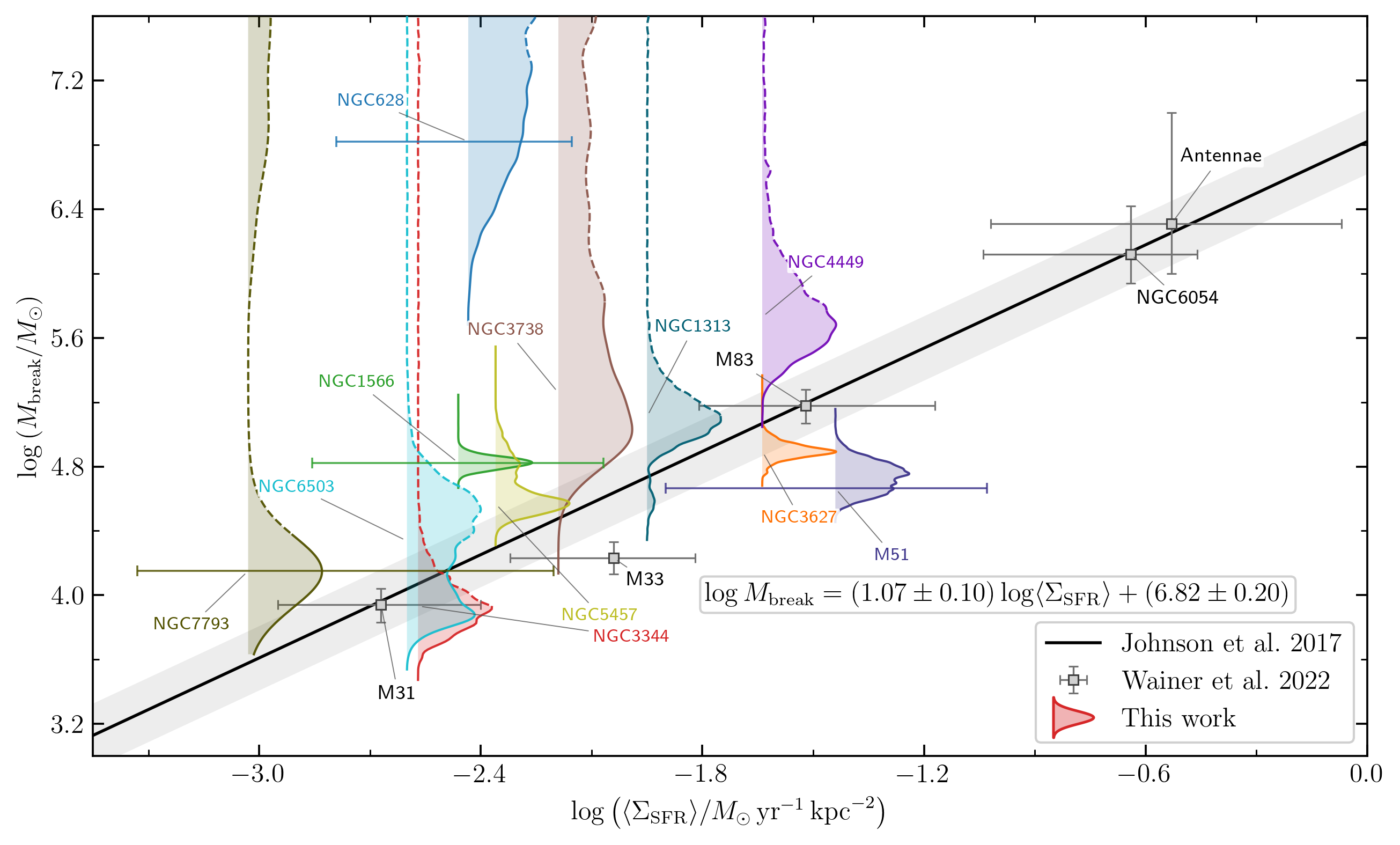}
    \caption{Posterior PDFs of $M_{\rm break}$ as a function of galaxy-wide $\Sigma_{\rm SFR}$. The black line shows the empirical relation from \citet{Johnson2017}, and the grey points show the literature measurements compiled by \citet{Wainer2022}. The coloured curves with shaded regions show the marginalized posterior PDFs for the galaxies of our work; dashed curves mark distributions with substantial high-$M_{\rm break}$ tails. Horizontal bars show the 68\% percent variation of local $\log\Sigma_{\rm SFR}$ measurements for those galaxies for which they are available. Galaxies for which no spatial range is reported are plotted without horizontal bars.}
    \label{fig:mid_mbreak_sigma_sfr}
\end{figure*}

\autoref{fig:mid_mbreak_sigma_sfr} compares our $M_{\rm break}$ posteriors with the empirical $\Sigma_{\rm SFR}$--$M_{\rm break}$ relation proposed by \citet{Johnson2017}, which builds on earlier work by several authors reporting evidence for a break in the otherwise-powerlaw CMF at high masses \citep[e.g.,][]{Bastian12b, Adamo2015, Adamo2017}. In this figure we plot our derived marginal posterior on $M_\mathrm{break}$ against the adopted representative galaxy-scale values of $\Sigma_{\rm SFR}$, using the values given in \autoref{tab:galaxy_properties}. For those galaxies where spatially resolved measurements are available and we can therefore quantify the extent of variation with position, we also provide error bars indicating this range -- see \autoref{sec:environmental_measurements} for details.

The figure shows that some of the galaxies in our sample show evidence for a finite break mass consistent with the proposed empirical relation (e.g., NGC~3344, 5457, and 6503, and NGC~5194--NGC~5195), but others have substantial probability mass at high $M_{\rm break}$ and remain consistent with the power law limit of the Schechter function (e.g., NGC 1313, 3738, and 7793), and some have posterior distributions of $M_\mathrm{break}$ that are strongly inconsistent with the proposed $\Sigma_\mathrm{SFR}-M_\mathrm{break}$ relation (NGC 628 and 1566). Our results are therefore not consistent with the existence of a simple relationship whereby galaxy-wide $\Sigma_{\rm SFR}$ uniquely determines the CMF break mass $M_{\rm break}$ in that galaxy. This finding is consistent with previous studies demonstrating that in at least some galaxies the CMF can be described as a pure powerlaw \citep{Cook2019,Mok2019,Mok2020}.

It is also worth noting that the bimodal marginal posterior PDFs for $M_\mathrm{break}$ shown by some of our sample (most prominently NGC 7793, but also visible for some other galaxies) arise because the parameters $\alpha_M$ and $M_{\rm break}$ are strongly covariant: a shallower asymptotic slope combined with a lower $M_{\rm break}$ can yield a mass function similar to that produced by a steeper slope and a much larger $M_{\rm break}$, at least over the range of mass accessible to our observations. In this situation a single, favoured value of $M_{\rm break}$ cannot be uniquely identified, which may explain some of the previous disagreement in the literature based on models that used simpler, non-Bayesian fitting methods -- if one were to attempt to derive $M_\mathrm{break}$ for NGC 7793 by simple $\chi^2$ fitting, the result could easily come out to either $\approx 10^{4.2}$ M$_\odot$ or $ > 10^{6.5}$ M$_\odot$ depending on details of the fitting method.

While this degeneracy can explain some of the scatter about the proposed \citeauthor{Johnson2017} relation that we find, it cannot be the sole explanation: our fit to NGC 628 excludes the break mass implied by this relation by more than 2 dex, and NGC 1566 by close to 1 dex. These deviations may arise because galaxy-wide $\Sigma_{\rm SFR}$ is only an indirect proxy for the physical quantities that actually regulate the maximum cluster mass. The relevant local conditions might include gas surface density, pressure, angular and epicyclic frequencies, and shear, which together might determine the maximum gas mass that can collapse before stellar feedback or galactic dynamics terminate star formation \citep{Kruijssen2012,ReinaCamposKruijssen2017}. A galaxy-wide catalogue may combine regions with very different values of this parameters, and thus very different break masses \citep{Adamo2015,Messa2018,Menon2021}. Spatial averaging may therefore broaden the inferred posterior. We therefore turn next to the implications of our findings on sub-galactic scales.


\subsection{Implications of sub-galactic-scale cluster demographics}
We find that the sub-galactic behaviour differs markedly among galaxies. NGC 5194 shows a clear radial decline in the inferred upper cluster mass scale, NGC 5457 shows a strong central-to-outer contrast, and NGC 628 shows essentially no change across the $\beta=0.3$ division, but does show strong variation between its central and eastern fields. We therefore ask whether these patterns can be understood within existing theoretical frameworks.


The motivation for dividing galaxies by $\beta$ stems from the shear-driven framework of \citet{SuwannajakTanLeroy2014}. In their model, star formation is regulated by cloud–cloud collisions driven by differential rotation. Regions with stronger shear (lower $\beta$, found in outer discs) experience higher collision rates, leading to enhanced star formation efficiency per orbital time, while higher $\beta$ corresponds to weaker shear and reduced collision rates (typically found in inner discs). Importantly, this model does not directly predict a dependence of the CMF truncation mass on $\beta$. Instead, any imprint on the CMF would arise indirectly: shear modifies the star formation efficiency, which could in turn influence the mass scale of star-forming structures and potentially propagate into the upper end of the CMF. Our subdivision therefore tests whether such large-scale kinematic differences leave any observable signature in the CMF, rather than testing a direct $M_{\rm break}$–$\beta$ relation.

A more general theoretical framework is provided by \citet[hereafter \citetalias{ReinaCamposKruijssen2017}]{ReinaCamposKruijssen2017}. The \citetalias{ReinaCamposKruijssen2017} framework links the maximum cluster mass to a combination of gravitational instability, galactic dynamics, and stellar feedback as a unified model. In this picture, the maximum cluster mass is set by the maximum mass of gravitationally bound giant molecular clouds and the efficiency with which these clouds are converted into bound clusters. This can be summarised as:
\begin{align}
M_{\rm cl,max} &= \epsilon \Gamma M_{\rm GMC,max}.
\end{align}
The maximum cluster mass is therefore determined by the maximum cloud mass able to collapse, the star formation efficiency $\epsilon$, and the cluster formation efficiency $\Gamma$. The maximum cloud mass depends primarily on the gas surface density $\Sigma_{\rm g}$ and the galactic dynamics, through the epicyclic frequency $\kappa$ and angular velocity $\Omega$, which set the shear and dynamical stability that regulate cloud collapse, while $\Gamma$ depends on the local gas and dynamical environment. Together, these quantities describe how much gas can collapse, how efficiently it forms stars, and what fraction of those stars remain in bound clusters. 

Having introduced the theoretical models for how galactic dynamics and local environmental parameters set the upper cluster mass scale encoded in the CMF, we test two questions: whether lower-$\beta$ regions exhibit the enhanced high-mass CMF that might follow from the cloud-collision picture, and whether departures from this trend can be understood qualitatively through the competing gas and dynamical effects in the multivariate \citetalias{ReinaCamposKruijssen2017} framework. To see this, we begin by reviewing each of the galaxies analysed in \autoref{ssec:env_dependence} individually, since the findings vary from one galaxy to another.  

\begin{figure*}
    \centering
    \includegraphics[width=0.99\textwidth]{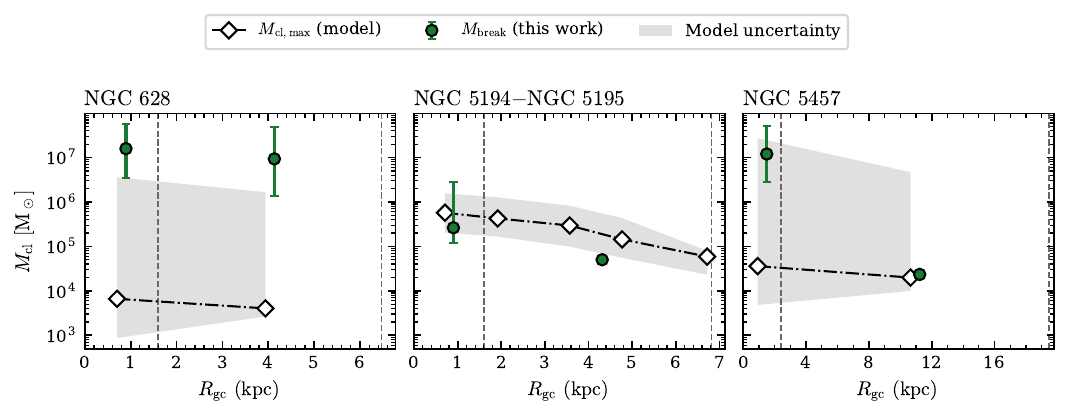}
    \caption{Comparison between the cluster mass-function break masses fitted in this study and the maximum mass scales predicted by the model of \citetalias{ReinaCamposKruijssen2017} for NGC~628 (left), NGC~5194--NGC~5195 (middle), and NGC~5457 (right), split into two radial bins at $R_{\rm gc}=1.6$ kpc for NGC~628 and NGC~5194--NGC~5195, and at $R_{\rm gc}=2.4$ kpc for NGC~5457. Vertical dashed lines mark the bin edges, and points are plotted at the bin midpoints; dash-dotted lines guide the eye. Black diamonds denote the maximum cluster mass scales predicted by the model. Green circles show $M_{\rm break}$ from this work, with asymmetric 16th--84th percentile intervals. The shaded regions show the 16th--84th percentile model uncertainty envelopes, incorporating the CO measurement uncertainties, the adopted $\alpha_{\rm CO}$ and $\kappa$ uncertainties, and the SFH correction -- see \aref{app:rck_figures}. For NGC~5194--NGC~5195, the model profile and uncertainty band are digitised from Fig.~18 of \citet{Messa2018}. Model and observed points are displaced slightly in galactocentric radius for clarity.
    }
    \label{fig:beta_628_compare}
\end{figure*}

\begin{figure*}
    \centering
    \includegraphics[width=0.9\textwidth]{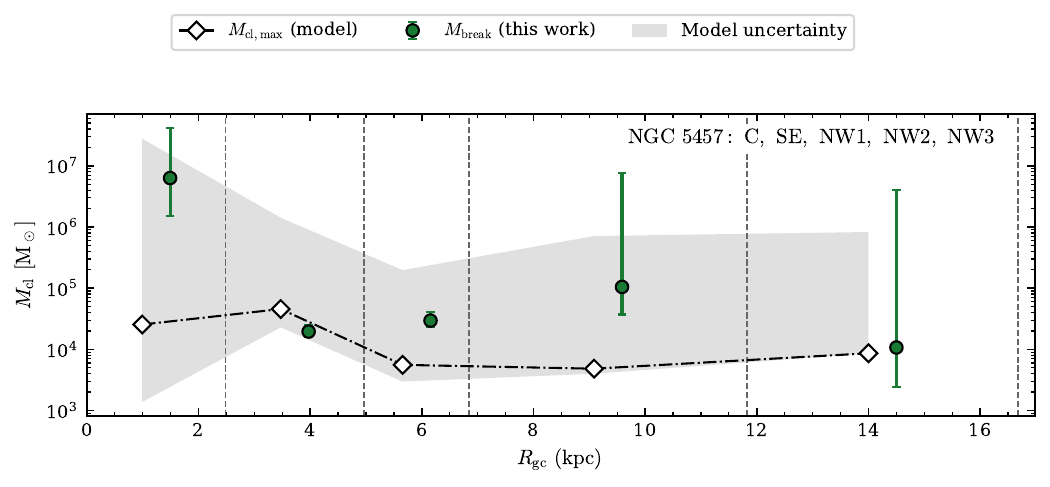}
    \caption{Same as \autoref{fig:beta_628_compare}, but grouped by the observational fields C, SE, NW1, NW2, and NW3 rather than radial bins. The vertical dashed lines mark the right-hand edges of the consecutive display intervals, and each field value is plotted at the corresponding interval midpoint.}
    \label{fig:fields_5457_compare}
\end{figure*}

\subsubsection{NGC~5457}
For NGC~5457, the 
central field is consistent with an effectively untruncated CMF, while the SE and NW1 fields have well-constrained, lower truncation masses. The NW2 and NW3 fields are less informative because of their smaller sample sizes, although they also provide tentative evidence for lower truncation masses as well. The division by $\beta$ tells a similar story, with the inner, high-$\beta$ region showing an effectively pure-powerlaw CMF while the outer, low-$\beta$ region shows clear evidence for a truncated CMF. This behaviour resembles the radial decline in CMF truncation mass observed in M83 \citep{Adamo2015}. 

These trends are in tension with the \citet{SuwannajakTanLeroy2014} framework, which predicts that lower-$\beta$, outer disc regions with stronger differential shear have higher star formation efficiencies per orbital time. If this higher efficiency were assumed to translate directly into a larger cluster formation efficiency and hence a higher truncation mass, the expected trend would be opposite to that observed: we would expect a higher CMF truncation mass in the high-shear, outer disc than the low-shear, inner disc. However, since their model predicts the star formation efficiency rather than the CMF truncation mass directly, this is only a point of tension rather than a direct conflict, and could be resolved if their model were to predict a lower CMF truncation mass in regions of higher star formation efficiency rather than a higher one as we have assumed.

To check the consistency of our findings with the \citetalias{ReinaCamposKruijssen2017} model we require three parameters: the gas surface density $\Sigma_{\rm gas}$, the epicyclic frequency $\kappa$ and velocity dispersion of gas $\sigma_{\rm g}$. We obtain these from the publicly-available THINGS H\,{\sc i} \citep{Walter2008}, HERACLES CO(2--1) \citep{Leroy2009}, and PHANGS-ALMA CO (2--1) \citep{Leroy2021} data sets, and use them to calculate the predicted maximum cluster mass $M_{\rm cl,max}$ in the \citetalias{ReinaCamposKruijssen2017} model, via the procedure described in \aref{app:rck_figures}. We compare this prediction to our measurements in the right panel of \autoref{fig:beta_628_compare}, which shows the results of dividing the galaxy radially based on where $\beta = 0.3$, and in \autoref{fig:fields_5457_compare}, which shows the results when we divide the galaxy by observational field; the black squares represent $M_{\rm cl,max}$, with the grey bands showing systematic uncertainties about these predictions computed as discussed in the Appendix, while green circles show our median $M_{\rm break}$ values with their 16th--84th percentile posterior intervals. 

The striking result from both of these figures is that we find a much sharper decline in $M_\mathrm{break}$ between the central region / field and the outer galaxy ones than the model appears to predict. While the very large uncertainty intervals on the model predictions (arising from the uncertain variation of gas properties in the past when the cluster population was forming -- see \aref{app:rck_figures}) means that the measurements are not explicitly outside the range of plausible model predictions, the model does not naturally reproduce such large variations with galactocentric radius.

\subsubsection{NGC~5194--NGC~5195}
For NGC~5194--NGC~5195, the inner, higher-$\beta$ region exhibits a more extended high-mass CMF and a higher inferred break mass than the outer region, as shown in \autoref{fig:beta_628_compare}. This trend is directionally consistent with the \citetalias{ReinaCamposKruijssen2017} predictions for NGC~5194 calculated by \citet{Messa2018}. In their model, the maximum cluster mass in the inner $\sim4\,{\rm kpc}$ is set by the Toomre mass, where shear and centrifugal forces limit the largest collapsing cloud. At larger radii, the limiting mechanism shifts to feedback limited collapse, causing the predicted upper cluster mass scale to decline. The predicted mass is a few $10^5\,M_\odot$ in the inner disc, while the feedback-limited branch reaches a few $10^4\,M_\odot$ in the outer disc. This decline is consistent with the break masses reported in \autoref{Tab:regional_cmf_params}, which decrease by approximately one dex from $\log_{10}(M_{\rm break}/M_\odot)=5.42$ in the $R_{\rm gal}<1.6\,{\rm kpc}$ region to $4.70$ at larger radii. However, our $\beta=0.3$ division at $1.6\,{\rm kpc}$ does not replicate the radial binning scheme used by \citet{Messa2018}. 

\subsubsection{NGC~628}
Applying the same division at $\beta=0.3$ to NGC~628 yields no significant difference in the CMF, as indicated by the strongly overlapping distributions in \autoref{fig:NGC628_beta}. This provides the clearest evidence in our sample against a universal one-parameter relation between $\beta$ and $M_{\rm break}$. Although the \citet{SuwannajakTanLeroy2014} model does not directly predict the CMF, the null result provides no support for a simple extension in which the higher star formation efficiency expected at lower $\beta$ produces a higher upper cluster mass scale. It may likewise challenge a simple dependence on gas surface density or pressure alone, if these quantities vary substantially across the division. 



To check whether the measurements are consistent with the \citetalias{ReinaCamposKruijssen2017} model, we again follow the procedure outlined in \aref{app:rck_figures} to compute the predicted maximum cluster mass $M_\mathrm{cl,max}$, yielding the black squares shown in the left panel of \autoref{fig:beta_628_compare}.
These predictions fall
substantially below the fitted $M_{\rm break}$ values; indeed, in these regions the $M_{\rm break}$ posteriors extend towards the upper boundary imposed by our priors, indicating that the CMFs follow effectively untruncated power laws. Even with the very generous uncertainty intervals on the theoretical predictions (again due to uncertainties of the star formation history -- see \aref{app:rck_figures}), the predictions do not reach such large masses.

\subsubsection{Summary}

Neither of the theoretical models we have tested provides a particularly convincing explanation of the trends in $M_\mathrm{break}$ revealed by our analysis. The shear-based model of \citet{SuwannajakTanLeroy2014} does not directly predict this quantity, but to the extent that we might expect it to be positively correlated with the star formation efficiency, which this model does predict, the predicted and observed trends run in opposite directions: the model predicts that star formation efficiency and thus likely maximum cluster mass is higher in the high-shear regions found in outer galaxies, while our analysis shows either the opposite trend (NGC 5194--5195 and NGC 5457) or no trend at all (NGC 628). The \citet{ReinaCamposKruijssen2017} model has a more complex dependence on gas conditions and shear, and applying that model predicts values of $M_\mathrm{cl,max}\sim 10^4 - 10^5$ M$_\odot$ with a mild declining trend with radius in NGC 628 and NGC 5457, and significantly larger masses $\sim 10^5 - 10^6$ M$_\odot$ but with a similar radial trend in NGC 5194--5195. Only the prediction in the last of these systems is plausibly consistent with the observations, and the model fails to reproduce the apparent lack of truncation of the mass function (i.e., the effectively pure powerlaw behaviour) we find in NGC 628 and in the central regions of NGC 5457.

\subsection{Systematic uncertainties in stellar population modelling}
Because our conclusions rely on forward modelling of the observed photometry, it is important to assess whether residual deficiencies in the stellar population models could bias the inferred mass-function parameters.

\begin{figure*}
    \centering
    \includegraphics[width=\textwidth]{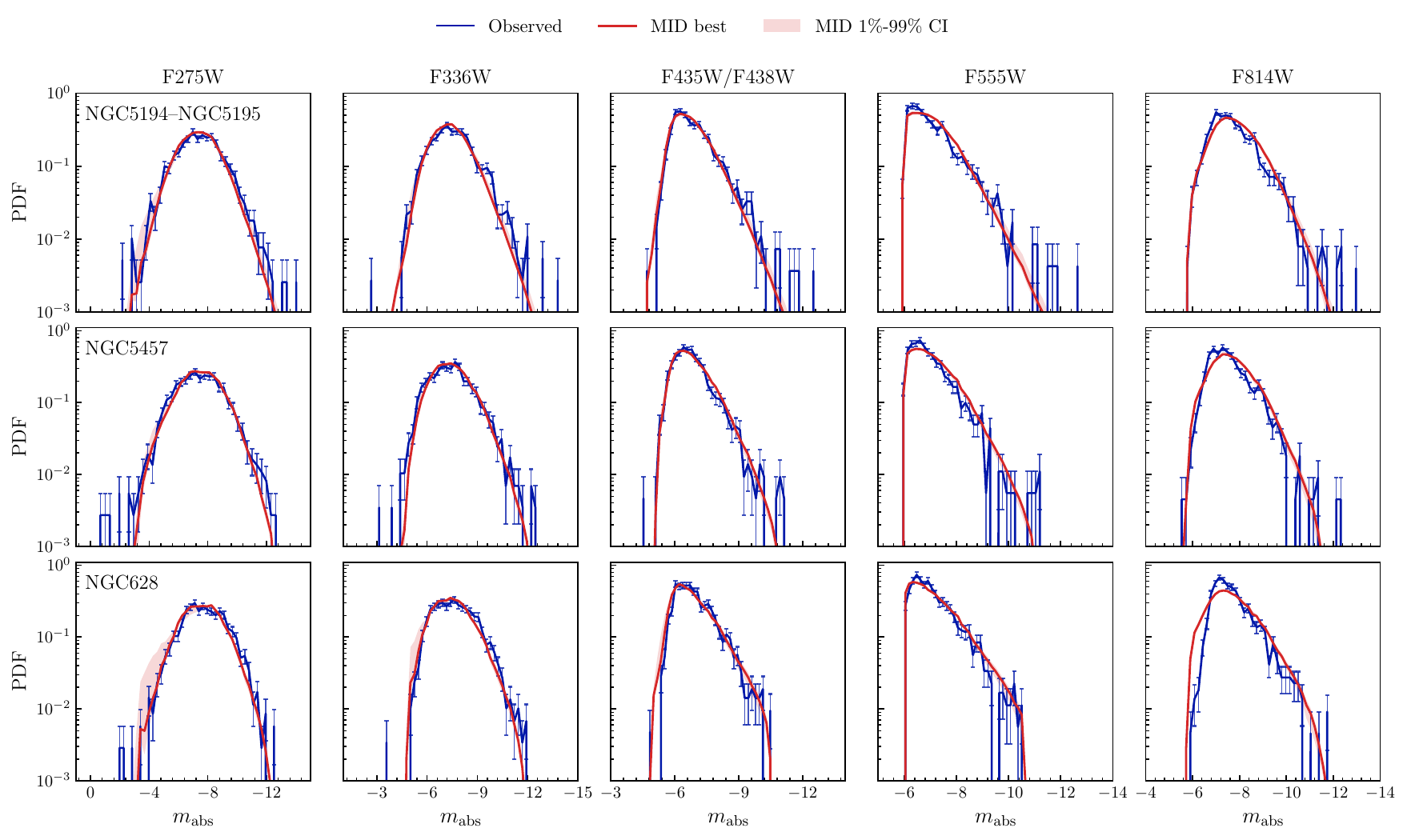}
    \caption{Comparison between the observed and forward-modelled absolute-magnitude PDFs for young clusters in NGC~5194--NGC~5195 (top), NGC~5457 (middle) and NGC~628 (bottom). The blue curves and error bars show the observed LEGUS catalogue distributions and their Poisson uncertainties, while the red curves show the MID model predictions obtained by reweighting the synthetic \slug~library according to the inferred demographic model and applying the catalogue selection function. The shaded regions indicate the central 1--99 percent intervals of the model realisations. Results are shown for the five LEGUS broad-band filters used in the fit. The model reproduces the overall UV--optical distributions well. The most prominent residual occurs in F814W for all three galaxies, where the observed distribution peaks at brighter magnitudes than the model prediction.}
    \label{fig:ngc628_luminosity_comparison}
\end{figure*}

The demographic inference method we use works by adjusting the parameters describing the star cluster population to make the theoretical forward-modelled photometric distribution match the observed distribution as closely as possible -- but this does not guarantee that even the best-fitting model provides a good representation of the data. Systematic deviations between data and model could arise both due to limitations of the functional forms we adopt to parametrise the cluster population (e.g., our functional forms for the CMF) and due to the limitations of the underlying stellar populations synthesis models (e.g., the stellar track and atmosphere library). It is therefore important not just to examine the best-fitting models, but also to assess how well those models are able to match the observed data.

To do so, we use our posterior PDFs to predict the luminosity function in each of our five bands. The full procedure is outlined in \citet{Tang2024}, but to summarise here: we randomly select a set of parameters $\boldsymbol{\theta}$ from our converged MCMC chains, and for each selection we calculate the statistical weight $w_{j,\mathcal{F}}(\boldsymbol{\theta})$ for each cluster $j$ in our synthetic library and for the field / filter set of interest $\mathcal{F}$ (\autoref{eq:library_weights}). We then compute the sum of $w_{j,\mathcal{F}}(\boldsymbol{\theta})$ for all library clusters in bins of photometric magnitude for each of our five bands, thereby predicting this photometric distribution function for that realisation. We repeat this procedure many times using different random samples from our converged chains, and find the median and 1\% to 99\% confidence interval in each bin. We then compare these predictions to the observed photometric distributions, for which we also generate Poisson uncertainty estimates via Monte Carlo resampling.

We show the outcome of this procedure in \autoref{fig:ngc628_luminosity_comparison} for three sample galaxies -- NGC~628, NGC~5457 and NGC~5194--NGC~5195, which we select as representative galaxies with large cluster samples. In \autoref{fig:ngc628_luminosity_comparison}, the blue curves show the observed LEGUS distributions in each filter, with error bars representing the Poisson uncertainties associated with the number of clusters in each magnitude bin. The solid red curves show the median predicted luminosity function, with the red shaded regions showing the central 1--99 percent intervals of Monte Carlo realisations of the fitted model. 

The observed and predicted distributions agree within the Poisson uncertainties in F275W, F336W, F435W, and F555W, indicating that the fitted mass--age--extinction distribution and the stellar models we use fully capture the principal UV and blue-optical photometric properties of the cluster population in all three selected galaxies. The model intervals generally broaden towards the faint end, where the expected number of clusters is smaller; this is expected behaviour, since our fitting procedure explicitly seeks to match not just the centre of the observed distribution, but also its uncertainties. The exception to this generally excellent agreement is in the F814W band, which shows a similar level of discrepancy for all three galaxies. In this band the observed distribution peaks at slightly brighter magnitudes than the synthetic-library prediction, and has a somewhat different shape. The discrepancy is therefore most likely band-dependent rather than galaxy-dependent. The residual is concentrated near the peak rather than being restricted to the faint, low-completeness regime, while the faint-end behaviour remains broadly consistent with the model. Its form is therefore not readily explained by an error confined to the catalogue completeness correction, although such a contribution cannot be excluded entirely. We instead suspect that the mismatch indicates the presence of a systematic associated with the stellar population synthesis modelling, rather than the catalogue-selection probability computed by \textsc{c-4}. 

One possible interpretation is that the stellar population synthesis models do not fully reproduce the reddest part of the spectral energy distribution of young clusters. At F814W, plausible sources of uncertainty include the treatment of short-lived luminous cool stars, stochastic occupation of red-supergiant and other evolved phases, nebular emission, and assumptions concerning attenuation and the extinction law \citep[e.g.][]{DaSilva2012,Krumholz2015,Hannon2019}. These effects cannot be separated using the present broad-band photometry, and the physical origin of the residual therefore remains uncertain. A possibly-related discrepancy has been reported at longer wavelengths by \citet{Pedrini2025}, who find that current stellar population models -- not just \slug, but other widely used models as well -- underpredict the $1.5$--$2.5\,\micron$ emission of young, low-mass emerging clusters observed with \textit{JWST}, even after accounting for stochastic IMF sampling, pre-main-sequence stars, and extinction. Additional contributions from circumstellar material, nebular emission, or hot dust may be required in that wavelength regime. It is unclear if the F814W residual identified here is related, since the wavelength regime is somewhat different, as are the sources targeted by LEGUS versus the JWST FEAST program on which the \citeauthor{Pedrini2025} analysis draws. Nevertheless, both results indicate that predictions at the reddest wavelengths are particularly sensitive to uncertain or missing ingredients in young-cluster population synthesis.

For the demographic analysis, the implication is limited but non-negligible. A band-dependent residual in the stellar population model can shift the inferred masses, ages, and extinctions of individual clusters through the degeneracies inherent to integrated-light photometry, and may therefore propagate into the recovered population-level mass and age distributions. We consequently treat the F814W mismatch as an additional systematic associated with the adopted stellar population model, distinct from uncertainty in the catalogue selection function. At the same time, the fact that the forward model reproduces the shapes, widths, and characteristic magnitudes of the observed one-dimensional distributions from F275W to F555W as well as possible given the limitations of the data supports our demographic inference at first order. The F814W residual is therefore best interpreted as a localized red-band modelling limitation rather than evidence for a broader failure of the inferred population model, although the present comparison does not establish its physical origin.

\section{Summary and conclusions}
\label{sec:summary}

A major challenge in star-cluster demographic studies is that observed cluster catalogues are strongly shaped by incompleteness, crowding, extinction, and catalogue-selection effects. To tackle this challenge, we have combined the stochastic stellar populations and Bayesian inference tool \slug~\citep{Krumholz2019SLUGIV} with the neural network-based completeness estimator \cfour~\citep{Tang2026} to infer star-cluster demographics from unresolved photometry in the LEGUS survey \citep{Calzetti2015, Adamo2017}. We use this approach to analyse \ngalaxy nearby galaxies spanning a broad range of morphologies, star-formation rate surface densities, and environments, comprising more than 8000 clusters, $\approx 30\%$ more than previous studies that were forced to apply strict cuts to the data to deal with poorly-known completeness.  Our completeness calculator implemented in \cfour~overcomes this problem by building a multidimensional forward model for star cluster catalogue inclusion, accounting for the dependence of recovery on cluster mass, age, extinction, photometry, effective radius, and local crowding without requiring binning or a single completeness threshold. This enables a homogeneous comparison of cluster-formation and disruption models across catalogues with different depths, filter coverage, and galactic environments. Our approach is able to reproduce the observations extremely well across the full galaxy sample, with predicted luminosity functions matching observed ones to within the Poisson uncertainties across all but the reddest of the five LEGUS bands, and the remaining discrepancies in the red plausibly explained by limits of stellar population synthesis models. 

We find that cluster mass functions (CMFs) are broadly consistent with an underlying power law slope $\alpha_M \approx -2$ over the mass range sampled by the data, but the evidence for a Schechter-like upper truncation above a finite characteristic mass $M_\mathrm{break}$ varies strongly between galaxies. We find clear support for a finite characteristic mass in NGC~1313, NGC~1566, NGC~3344, NGC~3627, NGC~5194--NGC~5195, NGC~5457, and NGC~6503, while NGC~628, NGC~3738, NGC~4449, NGC~5253 and NGC~7793 remain consistent with an effectively untruncated power law within the observed mass range, and for NGC 628 in particular the data strongly rule out any truncation mass below $\approx 10^6$ M$_\odot$. Marginal posterior distributions for $M_\mathrm{break}$ are often broad, skewed, or multimodal, reflecting a strong covariance between $\alpha_M$ and $M_{\rm break}$: a shallower low-mass slope with a lower break mass can reproduce nearly the same observed CMF as a steeper slope with a higher break mass. Consequently, the evidence for a truncation should be interpreted in terms of the full posterior distribution rather than a single best-fitting value of $M_{\rm break}$. Even within the broad uncertainties imposed by this covariance however, we find that proposed models for a relationship between $M_\mathrm{break}$ and galactic star-formation surface density $\Sigma_\mathrm{SFR}$ \citep[e.g.,][]{Johnson2017, Wainer2022} do not reproduce our observations; indeed, in our sample we find that $M_\mathrm{break}$ varies by $\gtrsim 2$ decades between galaxies whose star formation surface densities are all within a factor of two of one another.

We also find clear evidence for variation in the CMF on sub-galactic scales.
Dividing NGC~5457, NGC~5194--NGC~5195, and NGC~628 at a common rotation-curve slope, $\beta=0.3$, separates lower-shear inner regions from more differentially rotating outer discs and tests whether shear-related dynamics are associated with variations in the upper CMF. NGC~5457 and NGC~5194--NGC~5195 show clear inner--outer differences in CMF shape and truncation scale, whereas NGC~628 shows no comparable radial variation despite exhibiting differences between its individual LEGUS fields. Thus, neither $\beta$, galactocentric radius, nor galaxy-averaged $\Sigma_{\rm SFR}$ alone uniquely determines $M_{\rm break}$.

We also find substantial galaxy-to-galaxy variation in cluster age distributions. Our galaxy sample strongly favours a mass-independent disruption (MID) model whereby the age and mass distributions are separable over a mass-dependent disruption (MDD) model where they are not. Akaike weights favour MID over MDD by more than 90\% in 9 out of 12, with the other 3 showing weights near 50\%, indicating that the data are ambiguous. No galaxy significantly favours MDD. However, the parameters describing the disruption show extremely wide variation, with some galaxies showing age functions with clear declines at ages tens of Myr and others showing nearly flat age distributions out to almost a Gyr. 


Taken together, our results suggest that environmentally regulated cluster formation is more complex than can be captured by any single galaxy-scale parameter. The completeness-aware forward modelling weakens the evidence for a universal $\Sigma_{\rm SFR}$–$M_{\rm break}$ relation, while the sub-galactic analysis shows that $\beta$ alone is not sufficient to predict the upper cluster-mass scale. Comparison with the more complex model of \citet{ReinaCamposKruijssen2017} further indicates that present-day gas properties are not always sufficient, and that the temporal evolution of the ISM may play an important role. These results point towards a multivariate picture in which gas surface density, pressure, velocity dispersion, galactic dynamics, cluster formation efficiency, feedback, and the evolutionary history of the ISM all contribute to setting the upper cluster-mass scale. Future applications of this framework to extragalactic star cluster catalogues, together with spatially resolved molecular gas measurements and temporally informed environmental reconstructions, will provide a direct test of which local physical conditions regulate the upper end of the cluster mass function. The main result of this work is therefore not the rejection of environmental regulation, but the rejection of simple one-parameter environmental regulation.


\section*{Acknowledgments}
We thank Jonathan C. Tan for helpful suggestions. JT and KG are supported by the Australian Research Council (ARC) through the Discovery Early Career Researcher Award (DECRA) Fellowship (project no. DE220100766) funded by the Australian Government. MRK is supported by the ARC through Laureate Fellowship FL220100020. This research was supported by the National Computational Infrastructure (NCI), which is supported by the Australian Government, through the National Computational Merit Allocation Scheme and the ANU Merit Allocation Scheme (award jh2). This work is based on observations made with the NASA/ESA Hubble Space Telescope and obtained from the Space Telescope Science Institute, which is operated by the Association of Universities for Research in Astronomy, Inc., under NASA contract NAS 5–26555. These observations are associated with program No. 13364. 

\section*{Data Availability}
The cluster catalogue data underlying this study are publicly available through the Mikulski Archive for Space Telescopes (MAST). The software, analysis tools, dependencies, scripts, and data used to generate the figures are publicly available on \hyperlink{https://github.com/JianlingTang/tgkr2026b/tree/main}{Github}. The supplementary material provides the complete tabulated fitting results for the 12 galaxies and their subregions.



\bibliographystyle{mnras}
\bibliography{example} 

\appendix

\section{Evaluation of the RC\&K model}
\label{app:rck_figures}

Here we describe the procedure by which we generate the \citetalias{ReinaCamposKruijssen2017} model predictions shown in \autoref{fig:beta_628_compare} and \autoref{fig:fields_5457_compare}.

\subsection{NGC 628}

For NGC~628, we use the PHANGS--ALMA CO(2--1) moment-0, moment-1, and moment-2 products and the corresponding moment-0 and moment-2 uncertainty maps \citep{Leroy2021}; the THINGS H\,{\sc i} moment-0 map \citep{Walter2008}; the LEGUS footprint \citep{Calzetti2015,Adamo2017}; and the PHANGS rotation curve \citep{Lang2020}. Because the CO and H\,\textsc{i} maps have comparable angular resolutions of approximately 11 arcsec, we reproject the H\,\textsc{i} moment-0 map directly onto the PHANGS--ALMA CO grid without additional convolution.

Our first step is to extract gas surface densities from these data, which we do pixel by pixel over the intersection of the LEGUS coverage map and the set of pixels in the gas maps for which CO moments 0, 1, and 2 and H~\textsc{i} moment 0 are measured. We convert the moment 0 CO(2--1) intensities $I_{\rm CO(2-1)}$ to molecular gas surface density using
\begin{equation}
\Sigma_{\rm mol}
=
\frac{\alpha_{\rm CO(1-0)}}{R_{21}}
I_{\rm CO(2-1)}\cos i,
\label{eq:rck_sigma_mol}
\end{equation}
where $i = 35.1^\circ$ is the inclination \citep{Calzetti2015} and we adopt \citeauthor{Leroy2021}'s recommended values for the CO(2--1) to CO(1--0) and CO(1--0) to mass conversion factors, $R_{21} = 0.65$ and $\alpha_{\rm CO(1-0)}=4.35~
{\rm M_\odot\,pc^{-2}(K\,km\,s^{-1})^{-1}}$; this conversion includes helium. We similarly convert the H~\textsc{i} moment-0 maps to gas surface densities as
\begin{equation}
    \Sigma_{\rm atom} = 1.36 m_{\rm H} X_\mathrm{HI} I_\mathrm{HI} \cos i,
\end{equation}
where $X_\mathrm{HI} = 1.82\times 10^{18}$ cm$^{-2}\,\mathrm{(K\,km\,s}^{-1})^{-1}$ is the standard conversion factor from intensity to H~\textsc{i} column density in the optically thin limit, and the factor of 1.36 accounts for helium. We then sum to obtain our final gas surface densities $\Sigma_{\rm g}=\Sigma_{\rm mol}+\Sigma_{\rm atom}$.

We next compute the kinematic quantities
\begin{equation}
Q=\frac{\kappa\sigma_{\rm g}}{\pi G\Sigma_{\rm g}},
\qquad
\kappa=\sqrt{2(1+\beta)}\,\Omega,
\label{eq:rck_q}
\end{equation}
pixel-by-pixel as well. Here $\sigma_{\rm g}$ is the velocity dispersion and $\Omega=v_{\rm rot}/R$ and $\beta={\rm d}\ln v_{\rm rot}/{\rm d}\ln R$ are the angular velocity and logarithmic index of the rotation curve. We take $\sigma_{\rm g}$ from the CO(2--1) moment-2 map, and the rotation curve from \citet{Lang2020}. To obtain derivatives of the rotation curve, as required for $\beta$, we spline-smooth the published data. 

This supplies all the observational quantities required to evaluate the \citetalias{ReinaCamposKruijssen2017} model. Following their prescription, we first evaluate the Toomre- and feedback-limited GMC masses which are
\begin{equation}
M_{\rm GMC,T}
=
\frac{4\pi^5G^2\Sigma_{\rm g}^3}{\kappa^4},
\qquad
M_{\rm GMC,fb}
=
M_{\rm GMC,T}
\left(\frac{t_{\rm fb}}{t_{\rm ff,2D}}\right)^4,
\label{eq:rck_gmc_limits}
\end{equation}
where $t_{\rm ff,2D}=\sqrt{2\pi}/\kappa$. The operative maximum GMC mass is
\begin{equation}
M_{\rm GMC,max}
=
\min(M_{\rm GMC,T},M_{\rm GMC,fb}),
\label{eq:rck_gmc_max}
\end{equation}
because a cloud cannot exceed either the available unstable gas reservoir or the mass assembled before feedback halts collapse. We evaluate $t_{\rm fb}$ using the \citetalias{ReinaCamposKruijssen2017} prescription with $\phi_P=3$, $t_{\rm sn}=3$ Myr, $\epsilon_{\rm ff}=0.012$, and $\phi_{\rm fb}=0.16~{\rm cm^2\,s^{-3}}$.

The CFE at the feedback time, $\Gamma(t_{\rm fb})$, is calculated using the \citet{Kruijssen2012} model and its public Fortran implementation.\footnote{\url{https://github.com/mustang-project/CFE}} The calculator takes as input $\Sigma_{\rm g}$, $Q$, $\kappa/\sqrt{2}$, and $t_{\rm fb}$. The maximum cluster mass is then
\begin{equation}
M_{\rm cl,max}
=
\epsilon_{\rm cl}\Gamma(t_{\rm fb})M_{\rm GMC,max},
\label{eq:rck_cluster_mass}
\end{equation}
with $\epsilon_{\rm cl}=0.1$ \citep{ReinaCamposKruijssen2017}. We compute our final results for each radial bin or HST field from the pixel-level predictions as gas mass-weighted means,
\begin{equation}
\langle M_{\rm cl,max} \rangle_{\rm reg}
=
\frac{\sum_i\Sigma_{{\rm g},i} M_{{\rm cl,max},i}}
     {\sum_i\Sigma_{{\rm g},i}},
\label{eq:rck_weighted_mean}
\end{equation}
where the sum runs over all pixels in the region.

\subsection{NGC 5457}

Our procedure for NGC 5457 differs only slightly from that for NGC 628. Since NGC 5457 is not available in the public PHANGS-ALMA data we instead use the HERACLES CO(2--1) moment-0 map \citep{Leroy2009} to derive our molecular gas surface densities; we adopt $R_{21} = 0.70$ for this purpose, the value recommended by \citeauthor{Leroy2009}. We convolve the H\,{\sc i} products to the HERACLES resolution and reproject them onto the HERACLES grid to derive atomic gas surface densities in the same set of pixels as for the molecular gas; we adopt $i = 18^\circ$ \citep{Leroy2013} for the inclination.

To derive the kinematic quantities we require $\sigma_{\rm g}$ and the rotation curve. Because the HERACLES data do not provide moment-2, we take $\sigma_{\rm g}$ from the THINGS H~\textsc{i} moment-2 map instead. For the rotation curve, we adopt the parameterised functional form suggested by \citet{Leroy2013},
\begin{equation}
v_{\rm rot}(R)
=
v_{\rm flat}\left[1-\exp\left(-R/l_{\rm flat}\right)\right],
\qquad
\beta(R)=\frac{x}{\exp(x)-1},
\label{eq:ngc5457_rotation}
\end{equation}
where $x=R/l_{\rm flat}$, $v_{\rm flat}=210~{\rm km\,s^{-1}}$, and $l_{\rm flat}=1.2$ kpc. The remainder of the calculation follows exactly as for NGC 628.

\subsection{Uncertainty propagation}
\label{app:rck_uncertainties}

To estimate realistic uncertainties on the predictions that incorporate both measurement and systematic error, we generate 100 Monte Carlo realisations for each radial bin or observational field, which we then process through the same pipeline we use for the central values. The quantities that we vary in the Monte Carlo realisations are:
\begin{itemize}
    \item For each realisation, we draw CO(2--1) intensities in each pixel from a Gaussian distribution whose width is centred on the moment-0 map and with a dispersion given by the corresponding uncertainty map, with negative realisations set to zero. We do not apply a similar procedure to the H~\textsc{i} intensities because THINGS does not provide uncertainty maps.
    \item To account for systematic uncertainties in $\alpha_{\rm CO}$ when converting the intensities to surface densities, we also randomly draw a value of $\alpha_{\rm CO}$ for that realisation (using the same $\alpha_{\rm CO}$ for each pixel) from a lognormal distribution centred on our adopted values, and with a width of 0.28 dex for NGC~628 and 0.30 dex for NGC~5457 \citep{Sandstrom2013}.
    \item For NGC~628, we also generate pixel-by-pixel random realisations of $\sigma_{\rm g}$ by drawing from a Gaussian distribution whose width is set by the CO moment-2 uncertainty map. We do not carry out a similar procedure for NGC 5457 because THINGS does not provide an H\,{\sc i} moment-2 uncertainty map.
    \item For each realisation we multiply the central value of $\kappa$ for each pixel by a log-normal factor centred on unity with logarithmic dispersion $\log_{10}(1+f_\kappa)$. For NGC~628, we adopt $f_\kappa=0.168$ and 0.090 in the inner and outer bins. For the NGC~5457 radial bins, we adopt $f_\kappa=0.184$ and 0.149, and for the by-field analysis we adopt 0.171, 0.080, 0.093, 0.073, and 0.308 for C, SE, NW1, NW2, and NW3, respectively. We derive these $f_\kappa$ values differently for the two galaxies. For NGC~628, the PHANGS rotation-curve table provides asymmetric uncertainties on $v_{\rm rot}$ at each radius. We generate 1000 realisations of the tabulated rotation curve, refit the smoothing spline used for the central calculation, and recompute 
    \begin{equation}
        \beta=\frac{R}{v_{\rm rot}}\frac{{\rm d}v_{\rm rot}}{{\rm d}R},
        \qquad
        \kappa=\sqrt{2(1+\beta)}\,\frac{v_{\rm rot}}{R}.
    \end{equation} 
    For each realisation, we calculate the gas-mass-weighted mean $\kappa$ over the same valid pixels used for the model prediction, and then set $f_\kappa=(\kappa_{84}-\kappa_{16})/(2\kappa_{\rm fid})$, where $\kappa_p$ is the $p$th percentile value; this procedure gives 0.168 and 0.090 for the inner and outer bins divided by $\beta$, respectively. For NGC 5457, the analytic rotation curve we adopt has no associated uncertainty, and the HERACLES data provide no CO velocity-field uncertainty map. We therefore use the THINGS H\,{\sc i} moment-1 field to estimate the discrepancy between the observed gas kinematics and the analytic rotation curve. We fit the H\,{\sc i} rotation velocity in each region and define
    \begin{equation}
    \delta_v(R)=
    \frac{v_{\rm rot,HI}(R)-v_{\rm rot,model}(R)}
         {v_{\rm rot,model}(R)}.
    \end{equation}
    After interpolating this radial residual profile to the gas-map pixels,
    we calculate
    \begin{equation}
    f_{\kappa,{\rm region}}=
    \left[
    \frac{\sum_j\Sigma_{{\rm g},j}\delta_v(R_j)^2}
         {\sum_j\Sigma_{{\rm g},j}}
    \right]^{1/2}.
    \end{equation} For a fixed rotation-curve shape, $\kappa\propto v_{\rm rot}$ locally, so this quantity approximates the fractional uncertainty in the amplitude of $\kappa$. The NGC~5457 values are therefore empirical kinematic sensitivity terms, rather than formal $1\sigma$ uncertainties; they include offsets from the analytic curve due to non-circular motions or geometry.
    \item Finally, following the procedure recommended by \citetalias{ReinaCamposKruijssen2017}, for each Monte Carlo draw we evaluate $M_\mathrm{cl,max}$ both for the present-day, observed values of $\Sigma_{\rm g}$, and for values that are increased by a factor $f_\mathrm{hist}$ to account for the fact that the galaxy may have been more gas-rich when the clusters formed. Both evaluations use the same realisations of the measurement and associated uncertainties, so that they differ only in the assumed gas surface density. We adopt $f_{\rm hist}=4$ for C, SE, and NW1 and $f_{\rm hist}=10$ for NW2 and NW3. These choices are motivated by evidence for enhanced past star formation in the NGC~5457 outer disc \citep{Mihos2013,Mihos2018,Garner2022,Watkins2024}. For NGC~628, we adopt $f_{\rm hist}=10$ in both bins, motivated by the declining recent SFH inferred by \citet{Lomaeva2022}. 
\end{itemize}

Given our full set of Monte Carlo realisations, we generate the upper and lower envelopes of the uncertainty ranges we show in \autoref{fig:beta_628_compare} and \autoref{fig:fields_5457_compare} as
    \begin{align}
    M_{\rm cl,max}^{\rm low}
    &=
    P_{16}\!\left(M_{\rm cl,max}^{\rm current}\right),
    \\
    M_{\rm cl,max}^{\rm high}
    &=
    \max\!\left[
    P_{84}\!\left(M_{\rm cl,max}^{\rm current}\right),
    P_{84}\!\left(M_{\rm cl,max}^{\rm historical}\right)
    \right].
\end{align} 
Here $P_{16}$ and $P_{84}$ denote the $16^{\rm th}$ and $84^{\rm th}$ percentiles of the Monte Carlo realisations, and the superscripts ``current'' and ``historical'' correspond to evaluations using present day conditions or using conditions where the gas surface density was larger by a factor of $f_\mathrm{hist}$. The upper boundary therefore shows the higher of the two predictions from the present-day and historical ISM states. The envelope propagates the CO measurement uncertainties and the adopted uncertainties in $\alpha_{\rm CO}$, $\sigma_{\rm g}$ where available, and $\kappa$, together with the historical-ISM factor $f_{\rm hist}$. The grey region therefore combines the uncertainty in the present-day prediction with its sensitivity to the adopted historical ISM conditions.

\bsp	
\label{lastpage}
\end{document}